\documentclass{paper}
\usepackage[dvips]{color}
\usepackage{graphics,graphicx}
\usepackage{latexsym}
\usepackage{times}
\usepackage{natbib} 
\usepackage{amssymb,amsmath,mathrsfs}
\usepackage{amsbsy} 
\newcommand{\mG}{\mathcal{G}}
\newcommand{\R}{\Re} 
\newcommand{\tr}{\mathrm{tr}}
\def\mA{\mathcal{A}}
\newcommand{\diag}{\mathrm{diag}}
\newcommand{\tran}{\mathrm{T}}
\DeclareMathOperator*{\argmin}{arg\,min}
\newtheorem{theorem}{Theorem}[section]

\newtheorem{lemma}{Lemma}[section]

\begin{document}

\title{Lassoing Eigenvalues}

\author{David E. Tyler \thanks{Research supported in part by NSF Grant DMS-1407751.} 
 ~ {\large and} ~ Mengxi Yi \\
\normalsize{Department of Statistics \& Biostatistics} \\[-4pt]
\normalsize{Rutgers, The State University of New Jersey, Piscataway, NJ 08854, U.S.A.} \\[-4pt]
\normalsize{dtyler@rci.rutgers.edu; mengxi.yi@rutgers.edu}} 

\maketitle

\abstract{
The properties of penalized sample covariance matrices depend on the choice of the penalty function. 
In this paper, we introduce a class of non-smooth penalty functions for the sample covariance matrix, 
and demonstrate how this method results in a grouping of the estimated eigenvalues. We refer to this 
method as \emph{lassoing eigenvalues} or as the \emph{elasso}.
}


\section{Introduction and Motivation} \label{Sec:Intro}
Principal components play a central role in many multivariate statistical methods. 
In working with the sample principal component roots, i.e.\ the eigenvalues of the sample covariance matrix, it has long been recognized that the 
larger roots tend to be overestimated and the smaller roots tend to be underestimated. 
Consequently, numerous approaches have been 
proposed for shrinking eigenvalues together, e.g.\ bias-correction \citep{Anderson:1965}, decision theoretic \citep{Stein:1975, Haff:1991},  
Bayesian \citep{Haff:1980, Yang-Berger:1994}, and marginal likelihood \citep{Muirhead:1982}. 

The aim of this paper is to study penalization methods for shrinking eigenvalues towards each other, and in particular, 
penalization methods based on non-smooth penalties. The rationale for using non-smooth penalty functions is that such penalization
methods can not only shrink the eigenvalues towards each other, but can also result in partitioning the eigenvalues into sub-groups of equal eigenvalues, 
i.e.\  the eigenvalues are \emph{lassoed} together. 

Partitioning the principal component roots into distinct groups can be viewed as a model selection method, with each
of the $2^{q-1}$ possible partitions representing a different model. Models for which the $p < q$ smallest eigenvalues are taken to be equal are 
commonly referred to as sub-spherical models, factor models, reduced rank covariance models or spiked covariance models 
\citep{Anderson:2003, Baik:2006, Davis-etal:2014, Paul:2007, Johnstone:2001}. The more general case for which
different subsets of the eigenvalues are taken to be equal are sometimes referred to as multi-spiked or generalized spiked covariance models 
\citep{Bai-Yao:2012, Mestre:2008}. Some such models yield relatively sparse covariance models. An obvious example is the case for which all
the eigenvalues are taken to be equal, which corresponds to the covariance being proportional to the identity matrix. For this case, 
the $q(q+1)/2$ distinct elements of a covariance matrix of order $q$ is reduced to one parameter. 

The paper is laid out as follows. In section \ref{Sec:SCM}, regularized sample covariance matrices are first reviewed. Some general results on the
uniqueness and continuity of the path for penalized sample covariance matrices based on orthogonally invariant penalties, are then given in 
Theorem \ref{Thrm:nuniq}. Also, a relationship between penalized sample covariance matrices and the estimation of covariance matrices 
under constrains is established in Theorem \ref{Thrm:constrain}.
A class of nonsmooth penalties which has the effect of lassoing the eigenvalue together are introduced and treated in section \ref{Sec:non-smooth}. In particular, 
Theorem \ref{Thrm:nmin} gives a closed form for the solution path of the corresponding penalized sample covariance matrices. Section \ref{Sec:tuning} 
discusses selecting a penalty within this class of nonsmooth penalties, as well as selecting the tuning parameter for the penalty term via cross validation. 
Some asymptotic results are given in section \ref{Sec:asym}, wherein 
an application of the Mar\u{c}enko-Pasteur law is used to develop a promising choice for a penalty function. An illustrative example with discussion is 
presented in section \ref{Sec:discuss}. Proofs are given in an appendix.

\section{Regularized sample covariance matrices} \label{Sec:SCM}
Let $X = \{x_1, \ldots, x_n\}$ represent a $q$-dimensional sample of size $n$, with sample mean $\overline{x} = n^{-1} \sum^n_{i=1} x_i$ 
and sample covariance matrix $S_n = n^{-1} \sum^n_{i=1} (x_i -\overline{x})(x_i - \overline{x})^\tran$. When $S_n$ is nonsingular, which 
occurs with probability one for $n > q$ when random sampling from a continuous multivariate distribution, $(\overline{x}, S_n)$ uniquely minimizes 
\begin{equation} \label{eq:nlmu}  
l(\mu,\Sigma; X) = n\log\{ \det(\Sigma)\} + \sum_{i=1}^n (x_i - \mu)^\tran\Sigma^{-1}(x_i - \mu)
\end{equation}
over all $\mu \in \R^q$ and $\Sigma > 0$, i.e.\ the class of positive definite symmetric matrices of order $q$.
The function $l(\mu,\Sigma; X)$ corresponds, up to an additive constant, to two times the negative log-likelihood function under random 
sampling from a multivariate normal distribution. For singular $S_n$, which always occurs for $n \le q+1$, 
the function $l(\mu,\Sigma; X)$ is not bounded below. Hereafter, unless state otherwise, it is presumed that $S_n$ is nonsingular.

Even when $n > q$, the sample covariance matrix is not very stable for small or even moderate values of $n$. Consequently,
regularized or penalized sample covariance matrices have been introduced \citep{Huang-etal:2006, Bickel-Levina:2008, Warton:2008}. 
Since penalizing the covariance matrix does not effect the estimate for $\mu$, let
\begin{equation} \label{eq:nl}  l(\Sigma;S_n) = n^{-1} l(\overline{x},\Sigma; X)= \tr(\Sigma^{-1}S_n) + \log\{ \det(\Sigma)\}, \end{equation}
which is uniquely minimized over $\Sigma$ at $S_n$.
A penalized sample covariance matrix, say $\widehat{\Sigma}_{\eta}$, is then defined as a minimizer over $\Sigma > 0$ of the penalized
objective function
\begin{equation} \label{eq:nL} L(\Sigma;S_n;\eta) = l(\Sigma;S_n) + \eta~\Pi(\Sigma). \end{equation}
Here $\Pi(\Sigma)$, defined on $\Sigma > 0$, denotes a nonnegative penalty function, with $\eta \ge 0$ being a tuning constant. 
Since the function $l(\Sigma;S_n)$ is strictly convex in $\Sigma^{-1}$, so is $L(\Sigma;S_n,\eta)$ whenever the penalty function is convex in 
$\Sigma^{-1}$. In this case the minimizer is uniquely defined, with $\widehat{\Sigma}_{\eta}$ being a continuous function of $\eta$. 

Penalty functions which are convex in $\Sigma^{-1}$ include $\Pi_{l1}(\Sigma) = \sum_{j=1}^q \sum_{k=1}^q | \{\Sigma^{-1}\}_{jk} |$,
which arises in the graphical lasso \citep{Friedman-Hastie:2008}, and
 $\Pi_{kl}(\Sigma) = \tr(\Sigma^{-1}) + \log\{ \det(\Sigma)\}$, which corresponds to the
Kullback-Liebler distance, under the multivariate normal distribution, between $\Sigma$ and $I_q$. The \citet{Ledoit-Wolf:2004} regularized 
sample covariance matrix, defined as $\widehat{\Sigma} = (1-\beta) S_n + \beta I_q$ with $0 \le \beta < 1$ being a tuning parameter, can be shown to 
minimize \eqref{eq:nL} when $\Pi(\Sigma) = \Pi_{kl}(\Sigma)$ and $\beta = \eta/(1+\eta)$.

The Ledoit-Wolf estimator pulls the sample covariance matrix towards the identity matrix, whereas the goal in this paper is 
to pull the sample covariance towards proportionality with the identity matrix, i.e.\ shrink the eigenvalues together. 
When using the penalized approach, shrinking eigenvalues towards each other without penalizing the
scale of the covariance matrix implies the use of a scale invariant penalty, i.e.\ $\Pi(\Sigma) = \Pi(\gamma \Sigma)$ for $\Sigma > 0$ and
$\gamma > 0$.  The only scale invariant penalty which is convex in $\Sigma^{-1}$ is a constant penalty. 
For penalties which are not convex in $\Sigma^{-1}$, the uniqueness of a solution to  \eqref{eq:nL} is not immediate, 
nor do convex optimization methods necessarily apply.

For shrinking eigenvalues towards each other, aside from scale invariance, one may desire that the penalty attains its minimum at 
any $\Sigma \propto I$, and that it be a function of $\Sigma$ only through its eigenvalues. The last property is equivalent to using an
orthogonally invariant penalty function, i.e. $\Pi(\Sigma) = \Pi(Q \Sigma Q^\tran)$ for any $Q \in \mathcal{O}(q)$, the group of orthogonal
matrices of order $q$.
\begin{lemma} \label{Lem:orth}  
The function $\Pi(\Sigma)$ is orthogonally invariant if and only if for some symmetric, i.e.\ permutation invariant, 
function  $\pi: \R^q \to \R$, $\Pi(\Sigma) = \pi\left(\log(\lambda_1), \ldots, \log(\lambda_q)\right)$, 
where \mbox{$\lambda_1 \ge \cdots \ge \lambda_q > 0$} are the ordered eigenvalues of $\Sigma$. 
\end{lemma}
Hereafter, the ordered eigenvalues of $\Sigma$ are denoted by $\lambda_1 \ge \cdots \ge \lambda_q > 0$, and 
the ordered eigenvalues of $S_n$ are denoted by $d_1 \ge \cdots \ge d_q > 0$. When using an orthogonally invariant penalty,
the optimization problem \eqref{eq:nL} reduces to an optimization problem on the eigenvalues.
\begin{lemma} \label{Lem:diag}  
Suppose $\Pi(\Sigma)$ is orthogonally invariant. Using the spectral value decomposition, express
$S_n = P_nD_nP_n^\tran$ with $P_n \in \mathcal{O}(q)$, and where $D_n = \diag\{d_1, \ldots, d_q\}$. Then 
\[L(\Sigma; S_n,\eta) \ge L(P_n \Lambda P_n^\tran,S_n,\eta),\] where $\Lambda = \diag\{\lambda_1, \ldots, \lambda_q \}$. 
\end{lemma}

Lemma \ref{Lem:diag} implies that the eigenvectors of the penalized sample covariance matrix are the same as
those of the sample covariance matrix, with the associated eigenvalues following the same ordering. Hence, given
an orthogonally invariant penalty, any solution to minimizing $\eqref{eq:nL}$ has the form
$\widehat{\Sigma}_{\eta} = P_n \widehat{\Lambda}_{n,\eta} P_n^\tran$, where 
$\widehat{\Lambda}_{n,\eta} = \diag\{\widehat{\lambda}_1, \ldots, \widehat{\lambda}_q  \}$
with the diagonal terms corresponding to a global minimizer,  over $\lambda_1 \ge \cdots \ge \lambda_q > 0$, of the function
\begin{equation} \label{eq:lambda}
 \mathcal{L}(\lambda; d,\eta) = \sum_{j=1}^q \{ d_j/\lambda_j + \log(\lambda_j) \} + \eta \ \pi\left(\log(\lambda_1), \ldots, \log(\lambda_q)\right). 
\end{equation}
Since $e^{-z}$ is strictly convex and $d_j > 0$, it follows that, for any $\eta \ge 0$, the function
\begin{equation} \label{eq:halpha} h(y;d,\eta) = \sum_{j=1}^q \{ d_j e^{-y_j} + y_j \} + \eta \ \pi(y_1, \ldots, y_q) \end{equation}
is strictly convex on $\R^q$ whenever $\pi:\R^q \to \R$ is convex, and hence it is strictly convex on
the convex set $\{ y \in \R^q \ | \ y_1 \ge \cdots  \ge y_q  \}$. Furthermore, $h(y;d,\eta) \to \infty$ whenever $\|y\|\to \infty$. 
These observations, along with Lemmas \ref{Lem:orth} and \ref{Lem:diag}, yield the following result.  
\begin{theorem} \label{Thrm:nuniq}
Suppose $\Pi(\Sigma)$ is orthogonally invariant, and $\pi: \R^q \to \R$, as defined in Lemma \ref{Lem:orth}, is convex.
Then the function $L(\Sigma;S_n,\eta)$ has a unique global minimum over $\Sigma > 0$, specifically  $\widehat{\Sigma}_{\eta} = P_n \widehat{\Lambda}_{n,\eta} P_n^\tran$
where $P_n \in \mathcal{O}(q)$ is defined as in Lemma \ref{Lem:diag} and $\widehat{\Lambda}_{n,\eta} = \diag\{\widehat{\lambda}_1, \ldots, \widehat{\lambda}_q  \}$
with the diagonal terms corresponding to the unique minimizer of \eqref{eq:lambda} over $\lambda_1 \ge \cdots \ge \lambda_q > 0$. Furthermore, 
$\widehat{\Lambda}_{n,\eta}$ and $\widehat{\Sigma}_{\eta}$ are continuous functions of $\eta \ge 0$. 
\end{theorem}

Examples of penalty functions which satisfies the conditions of Theorem \ref{Thrm:nuniq} include the Kullback-Liebler penalty, since its
corresponding function $\pi(y) = \sum_{j=1}^q \{e^{-y_j} + y_j\}$ is symmetric and convex. A scale invariant example
is the eccentricity penalty $\Pi_e (\Sigma)  = \log\{\overline{\lambda}_a/\overline{\lambda}_g\}$,
which corresponds to the log of the ratio of the arithmetic mean $\overline{\lambda}_a$ to the geometric mean $\overline{\lambda}_g$
of the eigenvalues of $\Sigma$. Its corresponding function $\pi(y) = \log(\sum_{j=1}^q e^{y_j} ) - \sum_{j=1}^q y_j/q - \log(q)$ 
is again symmetric and convex. 

A problem related to the penalized covariance problem is the estimation of the covariance matrix under constrains. In particular, the next theorem 
establishes the following duality between the penalized problem and a constrained estimation problem.
\begin{theorem} \label{Thrm:constrain}
Assume the conditions of Theorem \ref{Thrm:nuniq}, and define $\kappa_L = \inf\{\Pi(\Sigma) ~|~ \Sigma > 0 \}$ and $\kappa_U = \Pi(S_n)$.
For $\kappa_L < \kappa \le \kappa_U$, there exists a unique solution $\widetilde{\Sigma}_\kappa > 0$ to 
the problem $\arg\min \{ l(\Sigma;S_n) \ | \ \Sigma > 0, \Pi(\Sigma) \le \kappa\}$, with the solution $\widetilde{\Sigma}_\kappa$ being a continuous 
function of $\kappa$. Furthermore, for each $\eta \ge 0$ there exists a $\kappa > 0$, and vice versa, 
such that $\widehat{\Sigma}_\eta = \widetilde{\Sigma}_\kappa$. The relationship between $\eta$ and $\kappa$ is
given by $\kappa(\eta) = \Pi(\widehat{\Sigma}_{\eta})$.
\end{theorem}

\section{Nonsmooth penalty functions} \label{Sec:non-smooth}
The choice of the penalty term $\Pi(\Sigma)$ and the tuning constant $\eta$ determines the nature and the extent to which the
eigenvalues are shrunk towards each other. In this section,
we study the following class of nonsmooth penalty functions which not only shrink the roots together, 
but generates equality for various subsets of eigenvalues for a large enough tuning constant.
\begin{equation} \label{eq:class}
\Pi(\Sigma;a) = \sum_{j=1}^q a_j \log(\lambda_j), \ \mbox{with} \ a_1 \ge \cdots \ge  a_q \ \mbox{and} \ \sum_{j=1}^q a_i = 0.
\end{equation} 
These penalty functions are scale and orthogonally invariant and, as the following lemma shows, satisfy the conditions of Lemma \ref{Thrm:nuniq}.
They are not differentiable in general since, although continuous, ordered eigenvalues are not differentiable functions at points of multiple roots.
\begin{lemma} \label{Lem:class}
For $a_1 \ge \ldots \ge a_q$, the function $\pi(y;a) = \sum_{j=1}^q a_j y_{(j)}$ is convex and symmetric, where $y_{(1)} \ge \ldots \ge y_{(q)}$
denotes the ordered values of $y \in \R^q$.
\end{lemma}
Note that $\Pi(\Sigma;a) = \pi(\log \lambda;a)$. If we had simply defined $\pi(y;a) = \sum_{j=1}^q a_i y_{j}$, then although it is convex, in particular linear,
it is not symmetric and so does not satisfy the conditions of Lemma \ref{Thrm:nuniq}.

The motivation for considering the class \eqref{eq:class} arose from first considering the special case
$\sum_{j<k} |\log(\lambda_j) - \log(\lambda_k)|$, which corresponds to choosing $a_1 = q-1, a_2 = q-3, \ldots,$
\mbox{$a_q = -(q-1)$}.
The absolute values signs in the penalty term is not, of course, necessary since $\lambda_j \ge \lambda_k$ for $j < k$, but are included to helps relate
the penalty to the $l_1$ penalty used in the regression lasso method. 
Other member of this class of penalty functions are discussed in the next section.

Using Theorem \ref{Thrm:nuniq}, finding the unique minimizer of $L(\Sigma;S_n,\eta)$ over $\Sigma > 0$ when using the penalty $\Pi(\Sigma;a)$
reduces to finding the unique minimizer of  
\begin{equation} \label{eq:lasso}
 \mathcal{L}(\lambda; d,\eta) = \sum_{j=1}^q \{ d_j/\lambda_j + (1 + \eta \ a_j)\log(\lambda_j) \} 
\end{equation}
subject to $\lambda_1 \ge \cdots \ge \lambda_q > 0$. To solve this optimization problems, first suppose the solution satisfies
$\lambda_1 > \cdots > \lambda_q > 0$, i.e.\ the minimum  occurs at a point where all the eigenvalues are distinct.  
In this case, the solution is simply the unique critical point $\widehat{\lambda}_j = d_j/(1+\eta a_j)$, \mbox{$j = 1, \ldots q$}. If this solution 
does not satisfy $\widehat{\lambda}_1 > \cdots \widehat{\lambda}_q > 0$, which will eventually occur with increasing $\eta$, then the true 
minimizer must contain at least one multiple root.

More generally, suppose the minimum of \eqref{eq:lasso} is achieved at a point where there are $r$ different eigenvalues of $\Sigma$,
say $\tilde{\lambda}_1 > \cdots > \tilde{\lambda}_r > 0$ with respective multiplicities $m_1, \ldots , m_r$, and hence $m_1 + \cdots + m_r = q$. 
Let $\mathcal{G} = \{ G(1), \ldots, G(r) \}$ denote the  corresponding partition of $\{1, \ldots, q \}$, i.e.\ 
$G(k) = (m_0 + \cdots + m_{k-1} + 1, \ldots, m_1 + \cdots + m_{k})$, with $m_0 = 0$. 
Given the presumed multiplicities, the objective function \eqref{eq:lasso} becomes
\begin{equation} \label{eq:lassoG}
\mathcal{L}_\mathcal{G}(\tilde{\lambda}; \tilde{d}) = 
\sum_{k=1}^r \{ \tilde{d}_k/\tilde{\lambda}_k + (1 + \eta \ \tilde{a}_k) \log(\tilde{\lambda}_k) \}, 
\end{equation}
where $\tilde{d}_k = \sum_{j \in G(k)} d_j/m_k$ and $\tilde{a}_k = \sum_{j \in G(k)} a_j/m_k$.
If $\mathcal{G}$ is the correct partition, then the minimum of \eqref{eq:lasso} is obtained at the unique critical point
of $\mathcal{L}_\mathcal{G}$, which is given by
\begin{equation} \label{eq:mult} \widehat{\lambda}_k(\mathcal{G}) = \tilde{d}_k/\{1 + \eta \ \tilde{a}_k \}, \ \mbox{for} \ k = 1, \ldots, r. \end{equation}
Again, if this solution does not satisfy  $\widehat{\lambda}_1(\mathcal{G}) > \cdots > \widehat{\lambda}_r(\mathcal{G}) > 0$, then the true minimizer 
does not correspond to the multiplicities implies by the partition $\mathcal{G}$. This implies the following condition on $\eta$.

\begin{lemma} \label{Lem:alphaG}
For $r>1$, 
the solution \eqref{eq:mult}, satisfies the constrain 
$\widehat{\lambda}_1(\mathcal{G}) > \cdots > \widehat{\lambda}_r(\mathcal{G}) > 0$ if and only if
$\eta <  \eta(\mathcal{G}) = \inf \{ \widetilde{\eta}_k(\mathcal{G}) \ | \ k = 1, \ldots r-1 \}$, where
\[ \widetilde{\eta}_k(\mathcal{G}) = \frac{\tilde{d}_k - \tilde{d}_{k+1}}{\tilde{a}_k \tilde{d}_{k+1} - \tilde{a}_{k+1}\tilde{d_{k}}} \]
provided $\tilde{a}_k \tilde{d}_{k+1} > \tilde{a}_{k+1}\tilde{d}_{k}$, and  $\widetilde{\eta}_k = \infty$ otherwise.
For $r = 1$, the solution $\widehat{\lambda}_1(\mathcal{G}) = \overline{d}$ holds for any $\eta <  \infty$.
\end{lemma}

The condition $\eta < \eta(\mathcal{G})$ is necessary but not a sufficient condition for $\mathcal{G}$ to be the correct partition.
It is possible for more than one partition to satisfy $\eta < \eta(\mathcal{G})$. In particular, it is always satisfied when $r=1$.
It remains then to find the correct partition $\mathcal{G}$.  For a given $\eta$, the minimizer of $\mathcal{L}(\lambda;d,\eta)$ must correspond to
the minimizer of $\mathcal{L}_\mathcal{G}(\tilde{\lambda}; \tilde{d})$ for some $\mathcal{G}$ for which $\eta < \eta(\mathcal{G})$, i.e.\
\[ \min_{\lambda_1 \ge \cdots \ge \lambda_q > 0}  \ \mathcal{L}(\lambda;d,\eta)  = 
\min_{\tilde{\lambda}_1 > \cdots > \tilde{\lambda}_r > 0} \  \{\mathcal{L}_\mathcal{G}(\tilde{\lambda};\tilde{d},\eta) ~|~ \eta(\mathcal{G}) > \eta\}. \]

It is not necessary to check all $2^{q-1}$ partitions of $\{ 1, \ldots, q \}$ to find the unique minimizer of  $\mathcal{L}(\lambda;d,\eta)$. Rather,
the unique minimizer can be found by considering only the following $q$ hierarchical partitions. Let $\mathcal{G}_q = \left\{ \{1\}, \ldots, \{q\} \right\}$.
For $\eta < \eta(\mathcal{G}_q)$, it readily follows that $\mathcal{G}_q$ is the minimizing partition. Next, define $\mathcal{G}_{q-1}$ to be the
partition formed by joining the two eigenvalues which become equal at $\eta = \eta(\mathcal{G}_q)$.
Continue in this fashion to produce the sequence  of partitions $\mathcal{G}_q, \ldots, \mathcal{G}_1$, with $\mathcal{G}_1 = \left\{ \{1, \ldots, q\} \right\}$. 
Specifically, given $\mathcal{G}_r = \{G_r(1), \ldots, G_r(r)\}$, define 
\begin{equation} \label{eq:Gr}
\mathcal{G}_{r-1} = \{ G_r(1), \ldots, G_r(k_r^*-1), G_r(k_r^*) \cup G_r(k_r^*+1), G_r(k_r^*+2), \ldots, G_r(r) \}, 
\end{equation}
where $k_r^* = \arg\inf\{k \ | \ \widetilde{\eta}_{k}(\mathcal{G}_r), k = 1, \ldots, r-1 \}$. Using this notation, we characterize the solution to minimizing 
$\mathcal{L}(\lambda;d,\eta)$ in the following theorem.  

\begin{theorem} \label{Thrm:nmin}
Suppose $d_1 > \cdots > d_q > 0$ and $k_r^*$, defined in \eqref{eq:Gr}, is unique for each $r = 2, \ldots, q$.
Then $0  < \eta(\mathcal{G}_q) < \cdots < \eta(\mathcal{G}_1) \equiv \infty$.
Furthermore, for $\eta(\mathcal{G}_{r+1}) \le \eta < \eta(\mathcal{G}_{r})$, with $\eta(\mathcal{G}_{q+1}) \equiv 0$,
$\mathcal{L}(\lambda;d,\eta) \ge \mathcal{L}(\widehat{\lambda};d,\eta)$, where $\widehat{\lambda}_j = \widehat{\lambda}_k(\mathcal{G}_r)$ for $j \in G_r(k)$.
Consequently, the unique minimizer to $L(\Sigma;S_n;\eta)$, when $\Pi(\Sigma) = \Pi(\Sigma;a)$, is given by 
$\widehat{\Sigma}_{\eta} = P_n\widehat{\Lambda}_{n,\eta}P_n^\tran$, where 
$\widehat{\Lambda}_{n,\eta} = \diag\{\widehat{\lambda}_1, \ldots, \widehat{\lambda}_q \}$.
\end{theorem}

The conditions in the preceding theorem hold with probability one when sampling from a continuous distribution. 
The conditions that $d_1 > \cdots > d_q > 0$ and $k_r^*$ be unique, though, are not necessary. They are included so that
the values of $\eta(\mathcal{G}_r)$ are all distinct. An extension of this theorem, which is needed in later sections,
is discussed in the appendix. 

When using $\Pi(\Sigma;a)$, the penalized method for estimating the covariance matrix is to be referred to as an \emph{elasso}. The elasso
has a number of properties similar to the lasso for regression. The estimated precision matrix $\widehat{\Sigma}_{\eta}^{-1}$ is a piecewise 
linear function of $\eta$, with the $q$ knots or kinks in the function occurring at $0 <\eta(\mathcal{G}_q) < \cdots < \eta(\mathcal{G}_2)$.  Hence, only
the knots and the values of $\widehat{\lambda}_1, \ldots, \widehat{\lambda}_q$ at the knots, together with $P_n$, need to be known to reconstruct 
the value of $\widehat{\Sigma}_{\eta}$ for all values of $\eta$. The value of the knots are easy to compute, and unlike the regression lasso,
the value of $\widehat{\lambda}$ at a knot has a simple closed form, namely it is a linear function of the sample eigenvalues. The knots of the
elasso yield a hierarchical set of $q$ models, namely $\mathcal{G}_q \succ \cdots \succ \mathcal{G}_1$, where $\mG_a \succ \mG_b$ implies
the sets in $\mG_b$ can be formed by unions of sets in $\mG_a$. In general, for $\eta(\mathcal{G}_{r+1}) \le \eta < \eta(\mathcal{G}_{r})$,
the grouping of the eigenvalues of $\widehat{\Sigma}_{\eta}$ consists of the $r$ groups indicated by the partition $\mathcal{G}_r$.

\begin{figure}[h]
\begin{center} 
\scalebox{.6}[.6]{
\vspace*{-2in}
\includegraphics{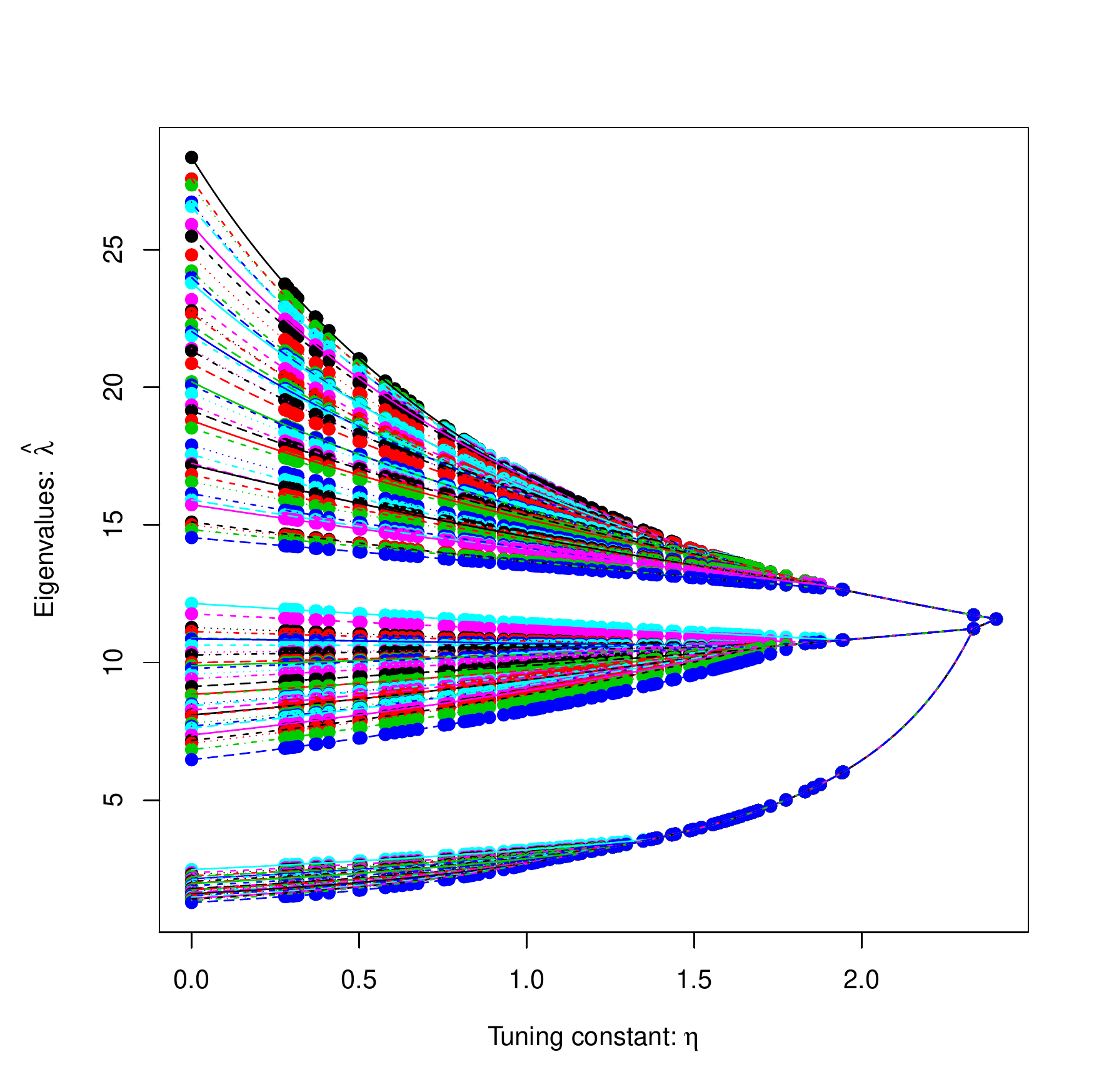}} 
\end{center}
\mbox{ } \\[-1.5cm]
\caption{An example of the elasso using Mar\u{c}enko-Pastur weights. \label{fig1} } 
\end{figure}

To illustrate the elasso, a pedagogical example is given in Figure \ref{fig1}, which shows the results from a simulated sample of size $n = 1000$ from a $q = 100$ 
dimensional multivariate normal distributions, for which the covariance matrix has $40$ eigenvalues equal to $20$, $30$ equal to $10$ and $30$ equal to $2$. 
The choice of the weights $a_1, \ldots, a_q$ used in the example are based upon the Mar\u{c}enko-Pastur law. These weights are discussed in the next section,
see \eqref{eq:mp}. The points displays in Figure \ref{fig1}
correspond to the knots where two eigenvalue groups are joined. Any eigenvalues that are joined at a given knot, remained joined for all $\eta$ greater
than that knot, hence producing the \emph{eigenvalue tree} and paths seen in the figure. The eigenvalue tree gives $100$ possible models or grouping of the eigenvalues,
and includes the true model at the third from the last knot, i.e.\ the model for which the multiplicities of the eigenvalues are $40, 30$ and $30$ respectively. 
\\

\section{Tuning the elasso} \label{Sec:tuning}
\subsection{Choice of weights} \label{Sec:wts}
In using the elasso, choices for the weights $a_1, \ldots, a_q$ and the tuning constant $\eta$ are needed.
The choice of weights partially depends upon the particular application of interest. Consider the condition
number penalty $\log(\lambda_1/\lambda_q)$, which corresponds to $a_1 = 1, a_2 = \cdots = a_{q-1} = 0$, 
and $a_q = -1$. This penalty lassoes only a group of the largest eigenvalues together and/or a group of the smallest eigenvalues 
together for any fixed $\eta$.  This follows by noting that for $1 < j < q$, one obtains $\widehat{\lambda}_j = d_j$ until $\widehat{\lambda}_j$
is joined to the largest or to the smallest eigenvalue group. As the value of $\eta$ increases, one eventually obtains a 
solution with only two distinct roots. Such a solution may be of interest if one is interested in producing a double spiked covariance model
with one spike representing the signal space and the other representing the noise space. The condition number has been considered by others 
for constraint likelihood problems \citep{Won-etal:2013, Wiesel:2012} but has not been previous studied as a penalty term. 

Another possible penalty term is
$\sum_{j=1}^q |\log(\lambda_j) - \log(\lambda_q)|$, for which $a_1 = \cdots = a_{q-1} =1$ and $a_q = -(q-1)$. This penalty lassoes only a 
group of the smallest eigenvalues together, i.e.\ any fixed $\eta$ yields a solution for which the smallest root having multiplicity $r$, 
with $r$ being an increasing function of $\eta$, and for which the $q-r$ larger roots having multiplicity one. This follows by noting that
since the weights $a_j, j = 1, ..., q-1$ are all equal to one, and so $\widehat{\lambda}_j$ and $\widehat{\lambda}_{j+1}$ cannot become equal 
at least until $\widehat{\lambda}_{j+1} = \widehat{\lambda}_q$. Lassoing only the smallest roots together can be used to obtain estimates, as
well as the rank of the signal space, in the single spiked covariance model or factor model. 

For detecting general multi-spike models, the weights $a_j$ should all be different. In this case, as $\eta$ increase,
the solutions behave in a manner similar to that displayed in Figure \ref{fig1}, i.e.\  two groups of roots come together at 
each knot until all roots are taken to be equal.
Based on some simulation studies, the penalty $\sum_{j<k} |\log(\lambda_j) - \log(\lambda_k)|$, previously discussed,
tends to keep the largest root separate, except for very large values of $\eta$, even when the largest 
population root is a multiple root. A more promising penalty is the one used in Figure \ref{fig1}. Here, the weights are obtained by centering 
decreasing quantiles from the \cite{MP:67} law, i.e.\ 
\begin{equation} \label{eq:mp} 
a_{mp,j} = \xi_j - \bar{\xi}, \ \mbox{where} \  \xi_j = F_{mp}^{-1}((q-j+.5)/q;q/n),
\end{equation}
with $F_{mp}(d;\nu)$ being the Mar\u{c}enko-Pastur cumulative distribution function with parameter $\nu$. Properties of the elasso based on this 
choice of weights are studied in section \ref{Sec:q/n}. 

\subsection{Using cross validation for choosing $\eta$ and for model selection} \label{Sec:CV}
There are a number of possible strategies for choosing the tuning parameter $\eta$. Here, we consider K-fold cross validation.  
For penalized approaches, cross validation can be applied to the unpenalized objective function, i.e.\ to \eqref{eq:nlmu} in this setting, which
gives
\begin{equation}
cv(\eta;\mA) = n_{\mA} \log\{\det(\widehat{\Sigma}_{-\mA.\eta})\} + 
\sum_{x_i \in \mA} (x_i - \overline{x}_{-\mA})^\tran \widehat{\Sigma}_{-\mA,\eta}^{-1} (x_i  - \overline{x}_{-\mA}), 
\end{equation}
calculated for a range of $\eta$ values \citep{Stone:1974, Huang-etal:2006}. Here $\mA$ denotes a subset of the data, with
$\overline{x}_{-\mA}$ and $\widehat{\Sigma}_{-\mA,\eta}$ representing, respectively, the sample mean vector and the penalized 
estimate of $\Sigma$ based on the data not in $\mA$. K-fold cross validation then seeks to minimize
$K^{-1}\sum_{k=1}^K cv(\eta;\mA_k)$ over $\eta \ge 0$, where $\mA_1, \ldots, \mA_K$ represents a random partition of the data into
subsets of equal size, plus or minus one.

\begin{figure}[h]
\begin{center}
\scalebox{.4}[.4]{
\includegraphics{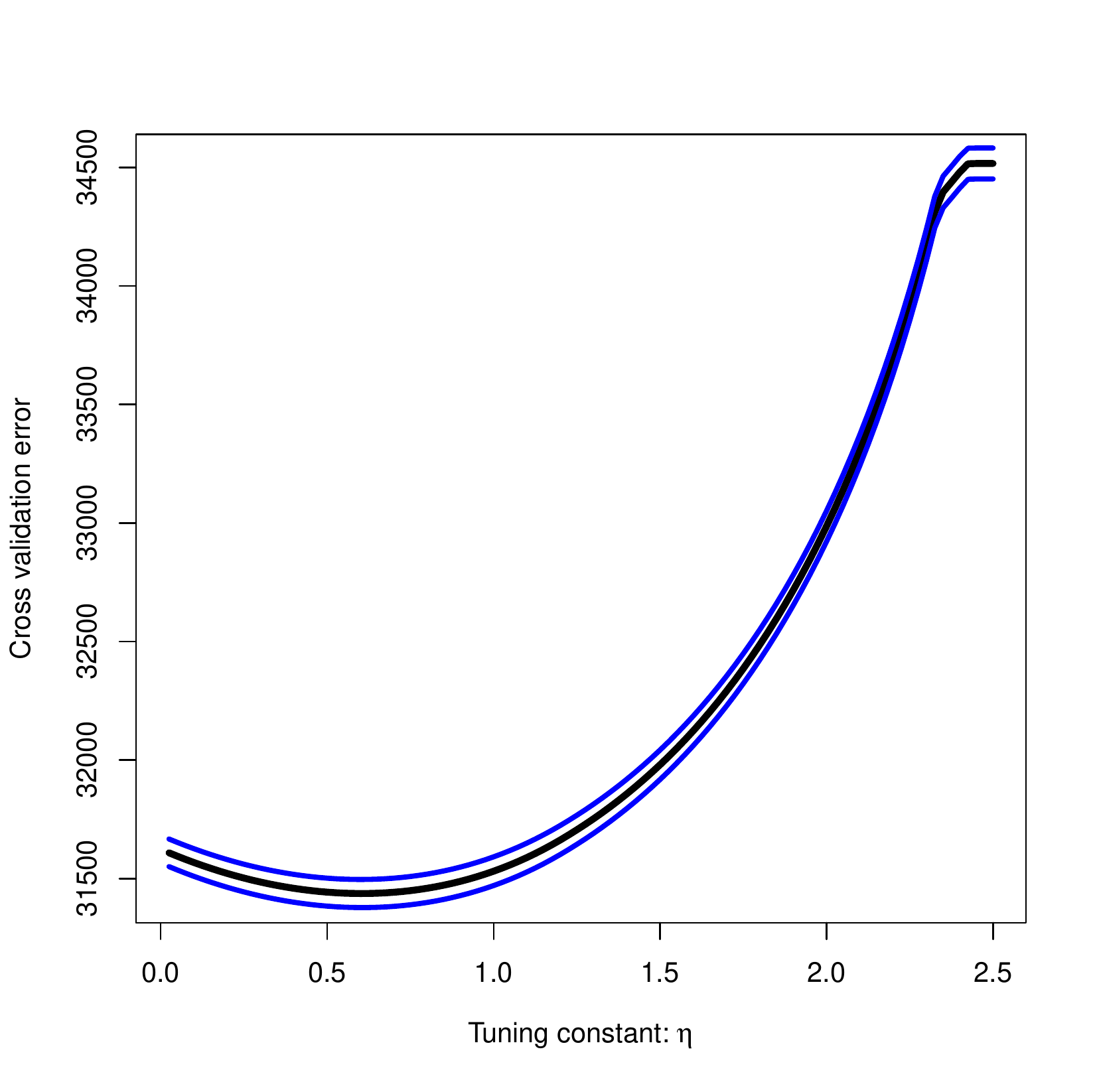}} 
\scalebox{.4}[.4]{
\hspace*{.1cm}
\includegraphics{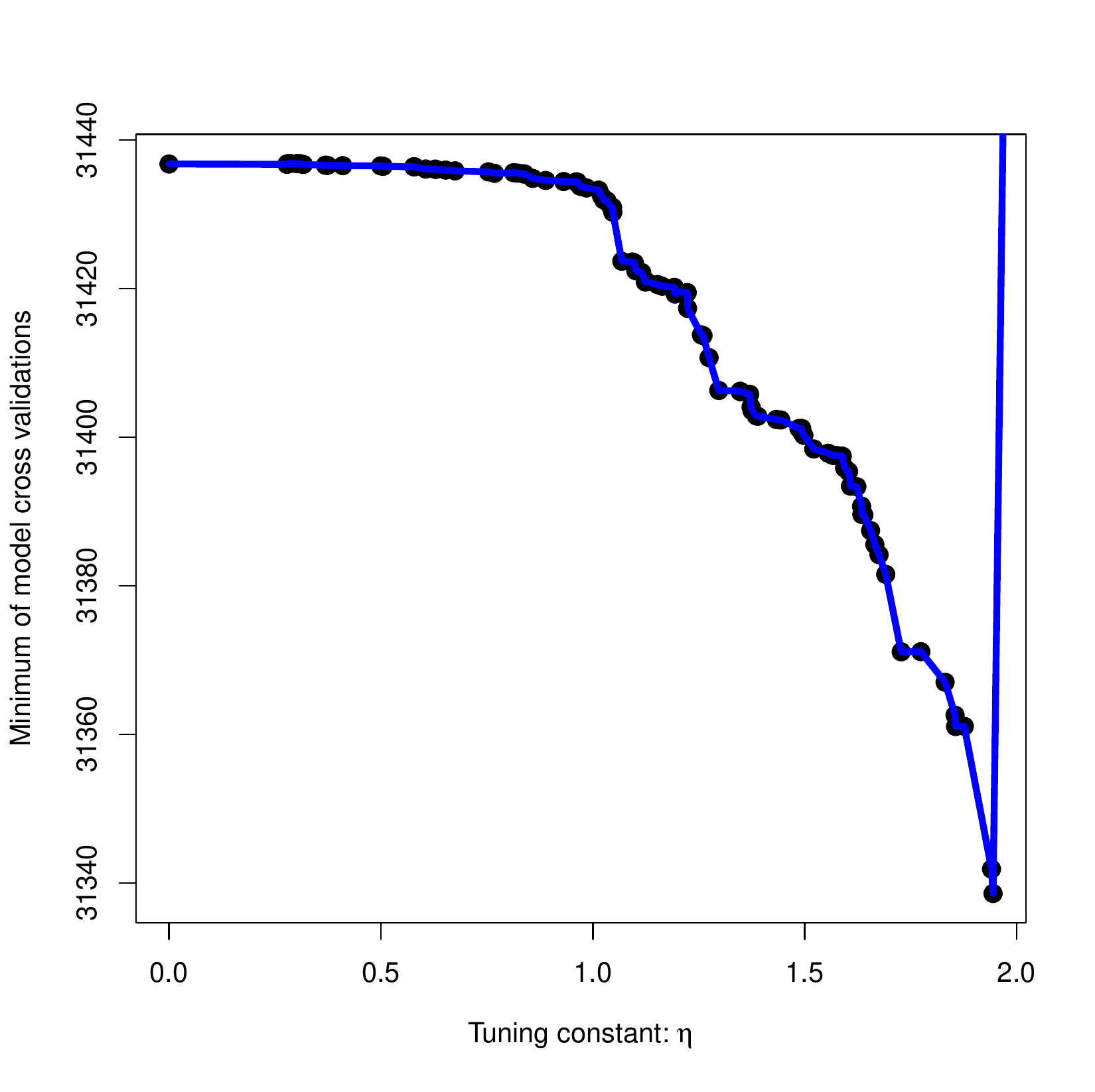}} 
\end{center}
\vspace*{-.6cm}
\caption{Ten fold cross validation (left) and model cross validation (right) results for the simulated data used in Figure \ref{fig1}. \label{fig2}}
\end{figure}

The graph on the left in Figure \ref{fig2} shows the results of a ten-fold cross validation for the 
data and weights used in Figure \ref{fig1}. The middle black curve corresponds to the mean of the $10$ values of 
$cv(\eta;\mA)$, with the blue lines corresponding to $\pm$ one standard error of the mean of these $10$ values. 
One hundred evenly spaced values between $0$ and $2.5$ are used for $\eta$. The minimum value 
in the plot is $31,436.78$, which is obtained at $\eta = 0.60$. Given the model used in the simulations,
it can be noted that the grouping of the eigenvalues in Figure \ref{fig1} at $\eta = 0.60$ is too coarse.

In regression lasso, a ``relaxed'' lasso is often recommended \citep{Mein:07} in order to obtain a simpler model. 
The analogy for the elasso would be to choose a larger value of $\eta$ having a cross validation mean equal to the cross 
validation $+$ one standard error at $\eta = 0.60$, which in this case corresponds to $\eta = 0.925$. 
Again, this does not yield a refined enough model. 

The reason why cross validation does not do well at selecting the correct model, which in this example
corresponds to three distinct roots with multiplicities $40, 30$ and $30$ respectively, is that the
correct model does not arise until $\eta = 1.95$. At this point, although the model is correct, the
roots are overly shrunk together and so the estimates of the parameters for this model result in a poor fit.

A proposed modification of the relaxed lasso is demonstrated in the right graph of Figure \ref{fig2}. Here,
ten-fold cross validation is performed at each of the $100$ models in the elasso path, with the 
minimum value of the cross validation for the model being plotted versus the corresponding model knot. 
It can be seen, in this case, that the model with the smallest cross validation error is the
correct multi-spiked covariance model. 

\begin{figure}[h]
\begin{center}
\scalebox{.40}[.40]{
\includegraphics{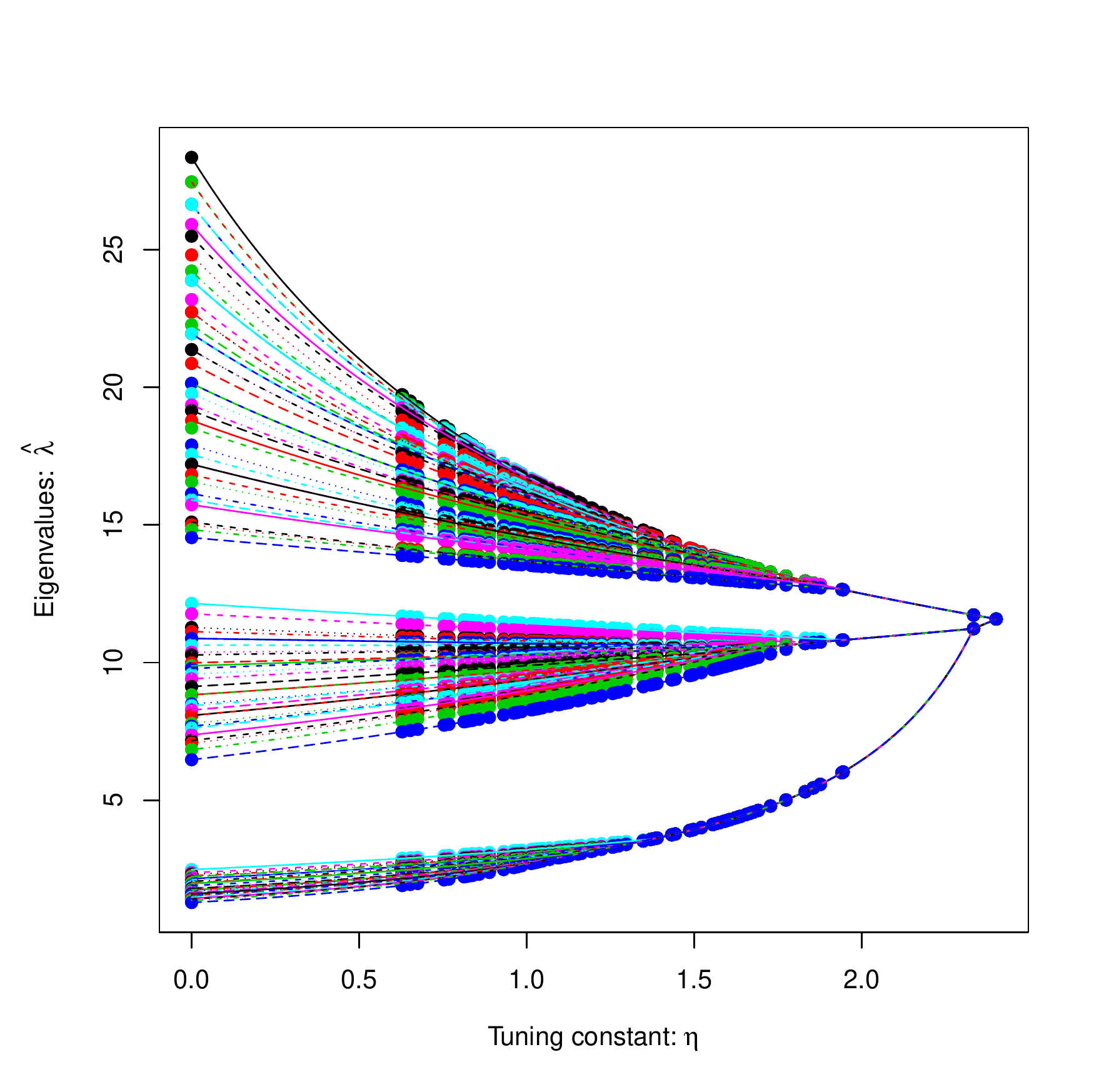}} 
\scalebox{.40}[.40]{
\hspace*{.1cm}
\includegraphics{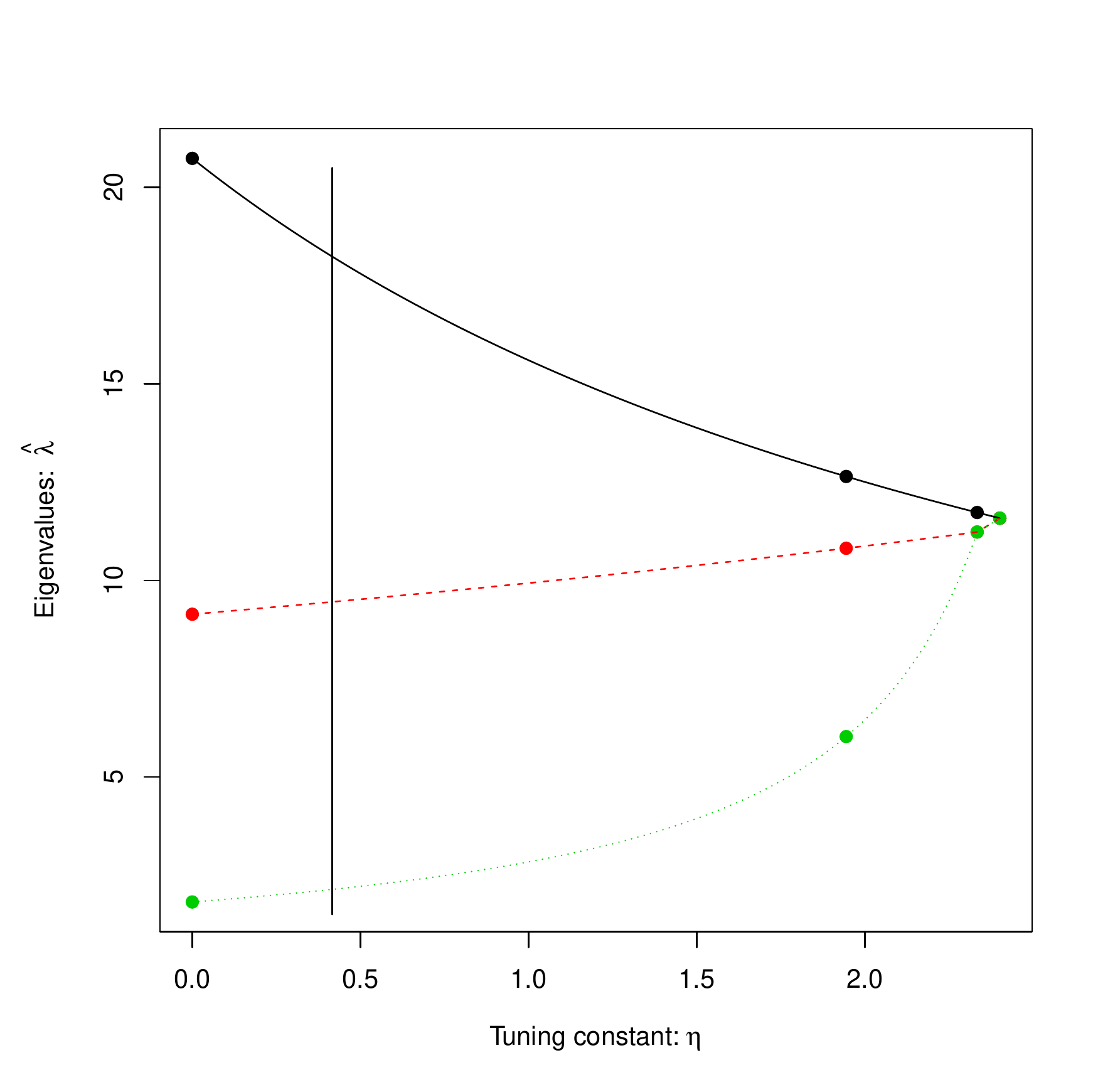}} 
\vspace*{-.6cm}
\end{center}
\caption{The elasso at selected multi-spike models. Left graph is for the model selected by cross validation, and the
right graph is for the model selected by the model cross validation. \label{fig3}}
\end{figure}

The elasso for a multi-spiked covariance model corresponding to the partition $\mathcal{G}_r = \{G(1), \ldots, G(r)\}$, as defined in \eqref{eq:Gr}, is obtained by
minimizing \eqref{eq:lassoG} over $\tilde{\lambda}_1 > \cdots > \tilde{\lambda}_r$. Theorem \ref{Thrm:nmin} readily extends to this case. The details are 
given in the appendix. For this case, the elasso path starts at $\eta = 0$ with the value of the estimated eigenvalues corresponding to 
their maximum likelihood estimate at the model, namely $\tilde{d}_1 > \cdots > \tilde{d}_r$, the average of the sample eigenvalues for each group of roots 
in the multi-spiked model. The elasso path is again linear in the inverse of the roots up to the knot for which the model arises, and then follows the 
original elasso path after the knot. Hence, the model elasso path has $r$ distinct knots. This is demonstrated in Figure \ref{fig3}. The left hand graph shows the 
elasso path corresponding to the multi-spike model given at $\eta = 0.60$. The right hand graph shows the multi-spike model selected via the model cross 
validation method, which in this case is the true model. The vertical line in the right hand graph is at $\eta = 0.42$, which is the value of $\eta$ 
producing the minimal cross validation error for this model. For this model, the cross validation means do not vary greatly for values of $\eta$ near
$0.42$, with the cross validation means for $0.05 < \eta < 0.79$ lying within one standard error of the minimal cross validation mean. 

The computational effort for model cross validation can be greatly reduced by only considering cross validation on the more parsimonious models, i.e.\
on models $\mG_r$ such that $\eta(\mG_r) > \eta_{min}$, where $\eta_{min}$ corresponds to the value of $\eta$ producing the minimum for 
the original cross validation, which in our example is $0.6$.
Also, for such $\mG_r$, we recommend that cross validation be performed only over the range $\eta < \eta_{\min}$.
For the model $\mG_r$ and $\eta \le \eta(\mG_{r-1})$, the inverse of the roots are linear in $\eta$, 
namely $\widehat{\lambda}_k(\mathcal{G}_r) = \tilde{d}_k/\{1 + \eta \ \tilde{a}_k \}$, for $k = 1, \ldots, r$. So,
for $\eta(\mG_r) > \eta_{min}$ and over the range $\eta < \eta_{\min}$, rather than perform an exhaustive cross validation on the model $\mG_r$,
we recommend performing cross validation on $\tilde{\Sigma}(\eta) = P_n\tilde{\Lambda}(\eta)P_n^{\tran}$, where
$\tilde{\Lambda}(\eta)$ is a diagonal matrix consisting of $\widehat{\lambda}_k(\mathcal{G}_r)$ repeated $r_k$ times for $k = 1, \ldots, r$.
This will yield the same result except in the unlikely case that one of the cross validation subsets yield an elasso for which the 
first knot after zero occurs before $\eta_{\min}$. In our example, performing this approximate cross validation only for models associated with 
knots greater than $0.60$ and only over the range $\eta < 0.60$ gives the same results as doing a complete model cross validation.

\section{Some Asymptotics} \label{Sec:asym}
\subsection{Fixed dimension} \label{Sec:fixed}
Although the focus of this paper is on introducing new methodology, some basic asymptotic justification can be given for the elasso.
The asymptotics as $n \to \infty$, with $q$ fixed, is relatively straightforward.  If the tuning parameter $\eta \to 0$ as $n \to \infty$, 
then the penalized estimator gives a consistent estimate. In particular, when $\eta$ is chosen by K-fold cross-validation for the elasso, then 
one obtains a consistent estimate of $\Sigma$. These assertions are stated formally in the following lemmas.

\begin{lemma} \label{Lem:asym1} 
Suppose $x_1, \ldots, x_n$ represents a random sample from $x$ having a $q$-dimensional distribution with mean $\mu_o$ and covariance matrix $\Sigma_o$.
Let $\widehat{\Sigma}_\eta$ be defined as in Theorem \ref{Thrm:nuniq}, with the conditions of the theorem holding. Then, \\[4pt]
 a) \ If $\eta \to 0$ as $n \to \infty$, then $\widehat{\Sigma}_{\eta} \to \Sigma_o$ almost surely. \\[4pt]
 b) \ Let $\eta^{cv}$ be the value of $\eta$ chosen via K-fold cross validation, i.e.\
\[ \eta^{cv}=\argmin_{\eta \ge 0} \left\{\sum_{k=1}^K cv(\eta;\mA_k)\right\},\] 
where $\mA_1, \ldots \mA_k$ is a random partition of $x_1, \ldots, x_n$ into subsets of equal size, plus or minus one.
For fixed $q$, as $n\to\infty$, $\widehat{\Sigma}_{\eta^{cv}} \to \Sigma_o$ almost surely.
\end{lemma}

The above lemma applies to any penalty function satisfying the conditions of Theorem \ref{Thrm:nuniq}, while the second lemma is stated only for
the elasso. An important feature of an elasso penalty, i.e.\ $\Pi(\Sigma;a)$ is that it generates knots. These knots are finite for any given data set, and
are random variables under random sampling. In particular, the last knot has the expression 
$\eta(\mG_1) = \sup\{ \eta_k \ | \ k = 1, \ldots q-1 \}$, where $\eta_k = (q D_k/D_q - k)/A_k$,
$D_k = \sum_{j=1}^k d_j$, and $A_k = \sum_{j=1}^k a_j$. Under the conditions of Lemma \ref{Lem:asym1}, if follows 
from the consistency of the sample eigenvalues that $\eta(\mG_1) \to 0$ almost surely whenever $\Sigma_o = \sigma^2 I$.
Hence, if the tuning parameter $\eta$ is chosen so that $P(\eta(\mG_1) < \eta) = 1 - \epsilon$ under spherical
multivariate normal sampling, then $\eta \to 0$. Such a choice for $\eta$ implies, not only is $\widehat{\Sigma}_\eta$
consistent for any $\Sigma_o$, but under multivariate normal sampling, the probability $\widehat{\Sigma}_\eta \propto I$ whenever 
$\Sigma_o \propto I$ is $1 - \epsilon$. Even under spherical normality, the distribution of $\eta(\mG_1)$ is complicated. However,
the distribution does not depend on the parameter $\sigma^2$, and so can be simulated using the standard normal distribution.  

The problem of identifying the correct multi-spike model needs further study.
As with cross validation, simulation studies imply choosing the tuning parameter $\eta$ in the above manner tends to yield
values of $\eta$ which are too small to identify the correct model whenever $\Sigma_o \not\propto I$.
Using cross-validation over the models, as described previously, though, appears to have a high probability of selecting the correct 
multi-spike model for large $n$.  Further work is needed to formally establish this assertion. However,
as stated in the following lemma, the elasso path is strongly consistent, i.e.\ the probability the path eventually contains the correct model 
as $n \to \infty$ is one.
\begin{lemma} \label{Lem:asymm}
Suppose the conditions of Lemma \ref{Lem:asym1} hold, with the multiplicities of the eigenvalues of $\Sigma_o$ corresponding to the
partition $\mG_o$. For the penalty $\Pi(\Sigma;a)$, define by \eqref{eq:class}, if $a_1 > \ldots > a_q$, then
\[P(\mG_o \in \{\mG_1, \ldots, \mG_q\} \ \mbox{for large enough n}) = 1. \]
\end{lemma}
Hence, of the $2^{q-1}$ possible multi-spike models, for large $n$ there is a high probability that the correct model is one of the
$q$ models in the path. Note that Lemma \ref{Lem:asymm} does not apply to the condition number penalty since in this case
$a_2 = \cdots = a_{q-1} = 0$.

\subsection{Increasing dimensions} \label{Sec:q/n}
When the sample size is small relative to the dimension, asymptotic approximations under the setting $n \to \infty$ with $q/n \to \nu \in [0,1)$
are of interest. A classical example is the \cite{MP:67} law, which states the following.  Suppose $x_1, \ldots, x_n$ represents a random sample from 
$x \in \R^q$, with $x$ itself having $q$ identical and independent components with unit variances and finite fourth moments. Let $F_n$ denote the 
empirical distribution of the eigenvalues $d_1 \ge \cdots \ge d_q$ of the sample covariance matrix $S_n$, i.e. $F_n(d) = \#\{d_i \le d \}/q$. 
Under this setting, $F_n(d) \to F_{mp}(d;\nu)$ almost surely, with $F_{mp}(d;\nu)$ being the Mar\u{c}enko-Pastur distribution with parameter $\nu$
and having density $f_{mp}(d;\nu) = (2 \pi x \nu)^{-1} \sqrt{(c_+-x)(x-c_-)}$ for $c_- \le x \le c_+$, where $c_\pm = (1 \pm \sqrt{\nu})^2$. 
 
A motivation for choosing the ``Mar\u{c}enko-Pastur'' weights, as described previously, in the elasso is the following. The two roots joined at the
first knot in the elasso are $d_{j^*}$ and $d_{j^*+1}$, where $j^*$ corresponds to the index for which the value of 
$\kappa_{j} = (d_j - d_{j+1})/(a_j d_{j+1} -a_{j+1}d_j)$ is minimized, but not negative, over $j = 1, \ldots, q-1$.  Under spherical normal sampling,
it would be desirable for $j^*$ to be purely random, i.e.\ uniform. Establishing such a result for given weights $a_1, \ldots, a_q$ appears to be rather
formidable. Alternatively, an approximate approach would be to choose weights so that the values of the random $\kappa_j$ are nearly equal. If one
could choose $a_j = \widehat{a}_j = (d_j - \overline{d})/\overline{d}$, then it readily follows that $\kappa_j = 1$ for $j = 1, \ldots, q$. 
However, $\widehat{a}_j$ is random rather than constant.  Under the asymptotic setting used to derived the Mar\u{c}enko-Pastur law, it follows that 
if $j/q \to p$ for some $p \in [0,1]$, then $\widehat{a}_j \to F_{mp}^{-1}(1-p;\nu) - 1 \to 0$ almost surely. Due to scale invariance of 
$\widehat{a}_j$, this limiting result also holds whenever the assumption of unit variance is replace by any variance $\sigma^2$. For finite
$n$ and $q$, the limiting value can be approximated by $a_{mp,j}$, which has the same limiting value as $\widehat{a}_j$.

\begin{figure}[h]
\begin{center}
\scalebox{.4}[.4]{
\includegraphics{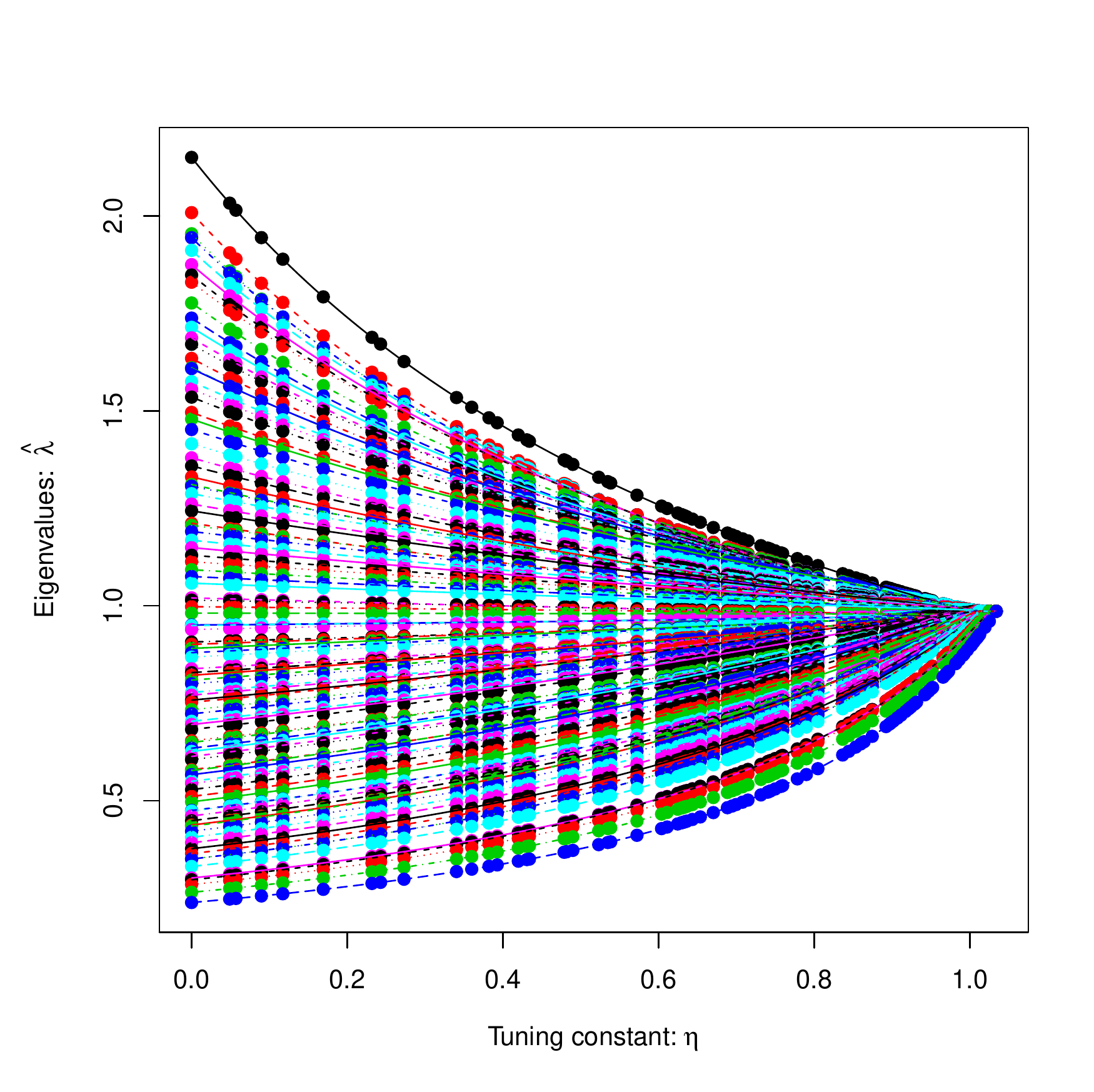}} 
\scalebox{.4}[.4]{
\hspace*{.1cm}
\includegraphics{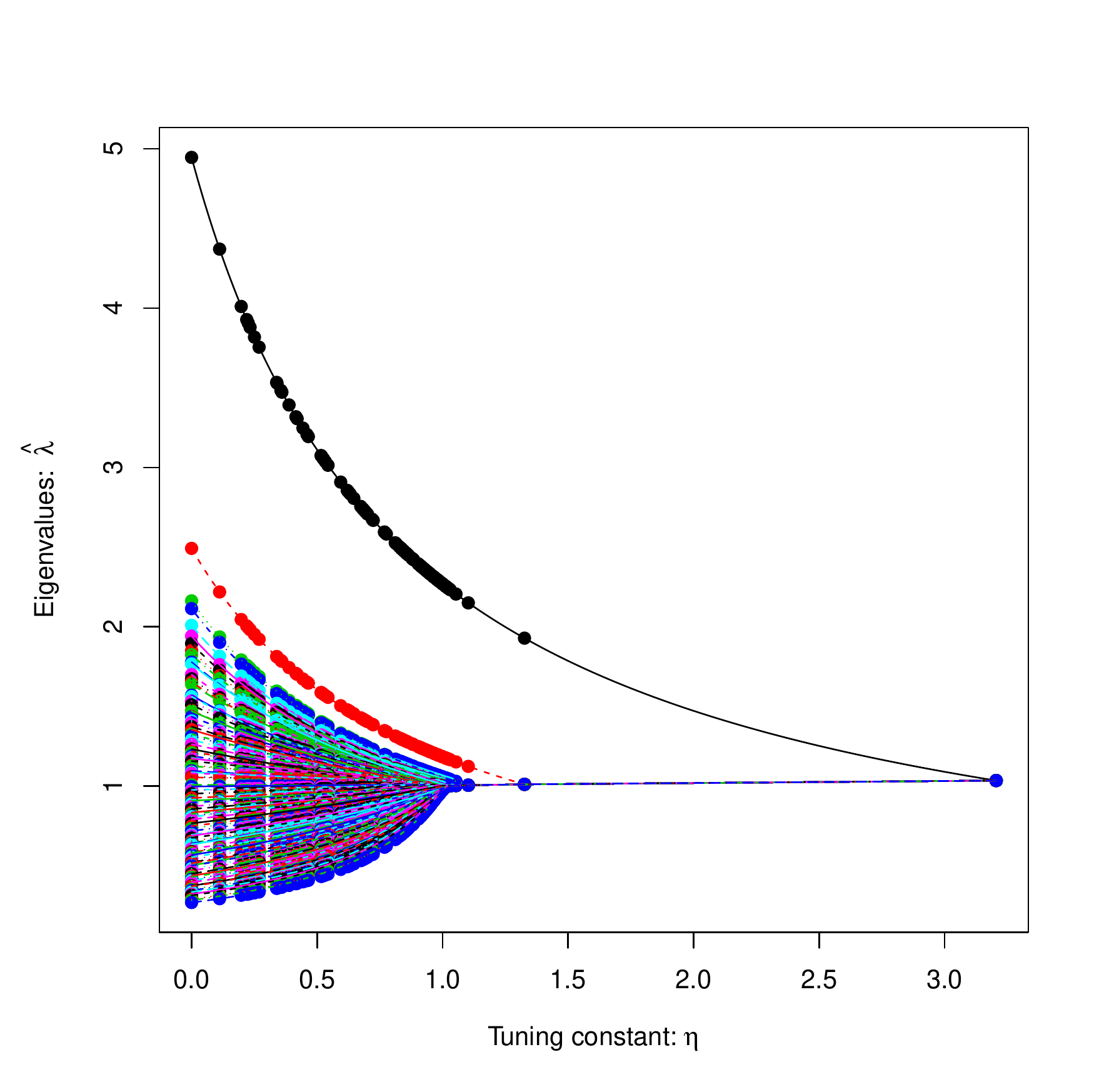}} 
\vspace*{-.6cm}
\end{center}
\caption{Examples of elassos using weights $a_{mp}$ for multivariate normal samples with $q=100$ and $n=400$. 
The eigenvalues of the covariance matrix are \mbox{$\lambda_1 = \cdots = \lambda_q = 1$ (left)}, and
$\lambda_1 = 4, \lambda_2 = 2$ and \mbox{$\lambda_3 = \cdots = \lambda_q = 1$ (right).} \label{fig4}}
\end{figure}

A formal study of the asymptotics in the large $n$ large $q$ setting is beyond the scope of the present paper.
However, some heuristic arguments, backed by simulation studies, suggests if $x \sim N_q(0,\sigma^2 I)$ and $j/q \to p, q/n \to \nu$ 
as $n \to \infty$, then the largest knot $\eta(\mG_{1}) \to 1$ almost surely when using the Mar\u{c}enko-Pastur weights in the elasso,
with the convergence to one tending to be from above. Also, for other knots $\eta(\mG_{j}) \to 1$ almost surely for any fixed $j$.
This is demonstrated in the plot on the left in Figure \ref{fig4} for the case $\sigma^2 =1$, $q = 100$ and $n = 400$.  
Consequently, if this conjecture holds and we choose $\eta > 1$, then the probability that the elasso solution  
$\widehat{\Sigma}_\eta = \overline{d} I$ goes to one. 
Choosing a fixed $\eta > 1$ does not imply inconsistency for the case $\nu = 0$, which
includes fixed $q$ as a special case. For the fixed $q$ case, the weights $a_{mp,j}$ depends on $n$, with $a_{mp,j} \to 0$ as
$n \to \infty$. So, if we standardized $a_{mp,j}^* = a_{mp,j}/a_{mp,1}$ and express $\eta~a_{mp,j} = \eta^* a_{mp,j}^*$, then for a fixed $\eta$, 
$\eta^* = \eta~a_{mp,1} \to 0$.

Results on the behavior of the knots when the true covariance matrix is not proportional to the identity can also be conjectured. 
An extension of the Mar\u{c}enko Pastur law  \citep{Baik:2006,Paul:2007} states the following. Suppose $x_1, \ldots, x_n$ represents a random sample from $x = Ay$,
$A \in \R^{q \times q}$ and $y \in \R^q$, with $y$ having $q$ identical and independent components with unit variances and finite fourth moments.
Also, suppose $\Sigma_o = AA^\tran$ has eigenvalues $\lambda_1 \ge \cdots \ge \lambda_{k} > \lambda_{k+1} = \cdots = \lambda_q = \sigma^2$. 
For $\lambda_{k} > \sigma^2(1+\sqrt{\nu})$, if $q/n \to \nu$ as $n \to \infty$, with $k$ held fixed,  
then $d_j \to \lambda_j^* = \lambda_j \{1+ \sigma^2\nu/(\lambda_j-\sigma^2)\}$ almost surely for $j = 1, \ldots, k$. Furthermore, 
the Mar\u{c}enko-Pastur law applies to the distribution of the $q-k$ standardized smallest roots, i.e. $d_j/\sigma^2$. 
This suggest that if $x \sim N_q(0,\Sigma_o)$, then $\eta(\mG_j) \to (\lambda_{j}^*/\sigma^2 - 1)/\{(1+\sqrt{\nu})^2 -1\}$ almost surely 
for $j = 1, \ldots, k$, whereas $\eta(\mG_j) \to 1$ for any fixed $j > k$.
This is demonstrated in the plot on the right in Figure \ref{fig4} for the case $\sigma^2 = 1$, $q = 100$, $n = 400$, $k=2$, $\lambda_1 = 4$ and $\lambda_2 = 2$.
If the last conjecture is true, then it implies the probability the smallest $q-k$ samples roots are grouped together in the elasso at $\eta = 1 + \epsilon$, 
but also remain separated from the largest $k$ sample roots, goes to one provided $\eta(\mG_k) > 1-\epsilon$. The last inequality holds if 
$\lambda_k > \sigma^2(1+\sqrt{\nu}) +\delta$, where $\delta \to 0$ as $\epsilon \to 0$. Consequently, under these conditions, the probability the elasso would correctly estimate $k$, the dimension of the signal space, goes to one. 

The extension of  Mar\u{c}enko Pastur law also states that if 
$\sigma^2 < \lambda_{j} \le \sigma^2(1+\sqrt{\nu})$, for some $j = 1, \ldots, k$, then $d_j/\sigma^2 \to (1+\sqrt{\nu})^2$, the maximum of the support 
of the Mar\u{c}enko-Pastur law. If $k^*$ is the smallest $j$ for which this condition on $\lambda_j$ holds, then the Mar\u{c}enko-Pastur law holds for the distribution of the $q-k^*+1$ standardized smallest roots. Under these conditions, it is not possible to consistently estimate $k$, 
but only $k^*$. The probability the elasso based upon the Mar\u{c}enko-Pastur weights estimates the dimension of the signal space to be $k^*$ then goes to one.

\section{An example and concluding remarks} \label{Sec:discuss}
\subsection{Telephone call centre data} \label{Sec:call}
As an example, we consider the call centre data previously analyzed by \citet{Huang-etal:2006} and \citet{Fan-etal:2009}. For each weekday in 2002, 
except for holidays and six days when the data collecting equipment was out, phone calls were recorded from 7:00 am until
midnight, resulting in a sample size of $239$. For each of these days, the responses correspond to the number of calls 
received in consecutive 10 minute periods, resulting in a $q = 102$ dimensional response vector $N$.  Since the number of calls
tend not to be normally distribution, each data point is then transform to $x = \sqrt{N + 0.25}$, where the operation refers to
each of the elements of $x$ and $N$. The sample $x_1, \ldots, x_{239}$ are presume to be independent observations.
A more complete description of the data can be found in \citet{Huang-etal:2006}. 

\begin{figure}[h]
\begin{center}
\scalebox{.4}[.4]{
\includegraphics{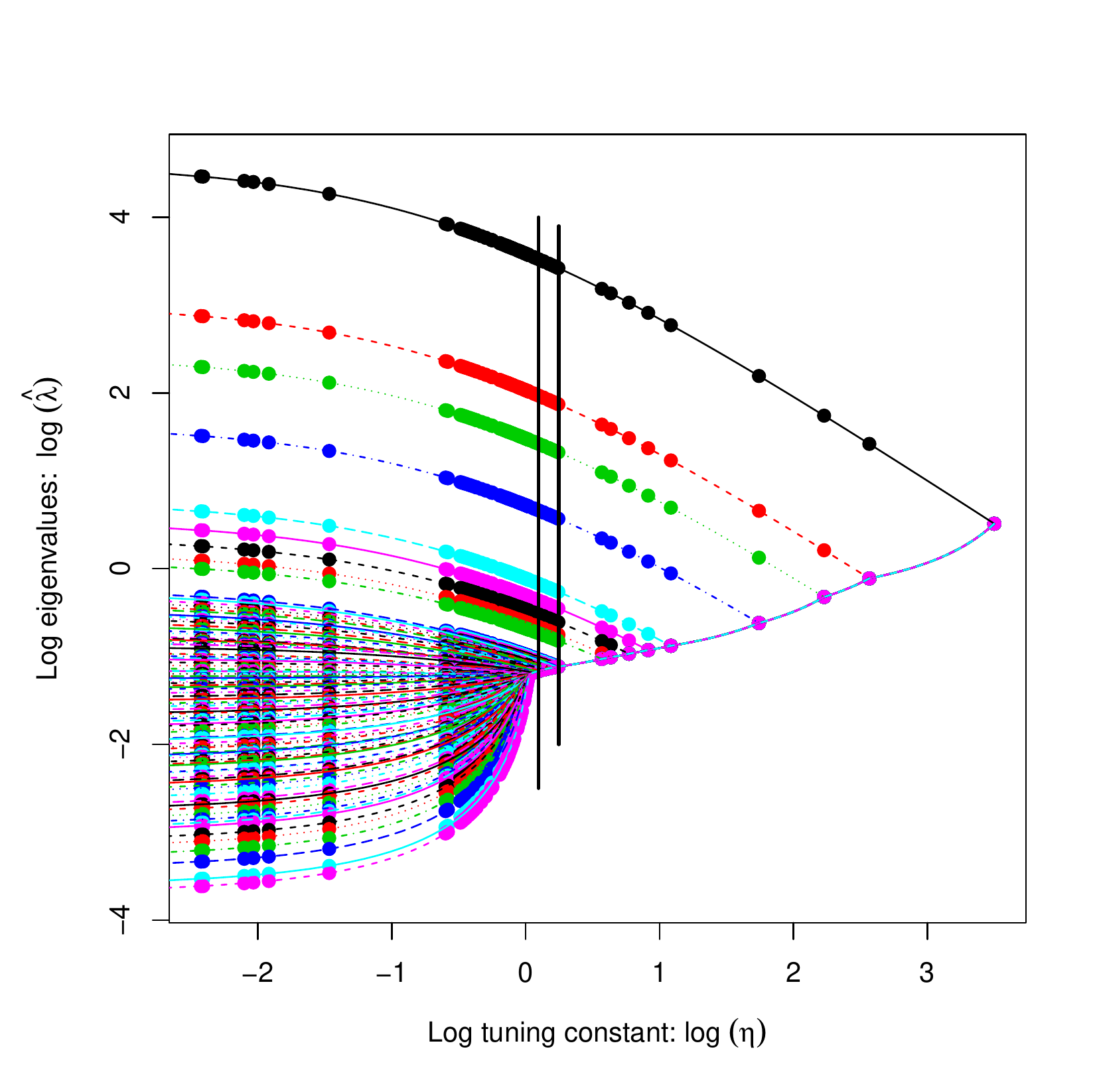}} 
\scalebox{.4}[.4]{
\hspace*{-.1cm}
\includegraphics{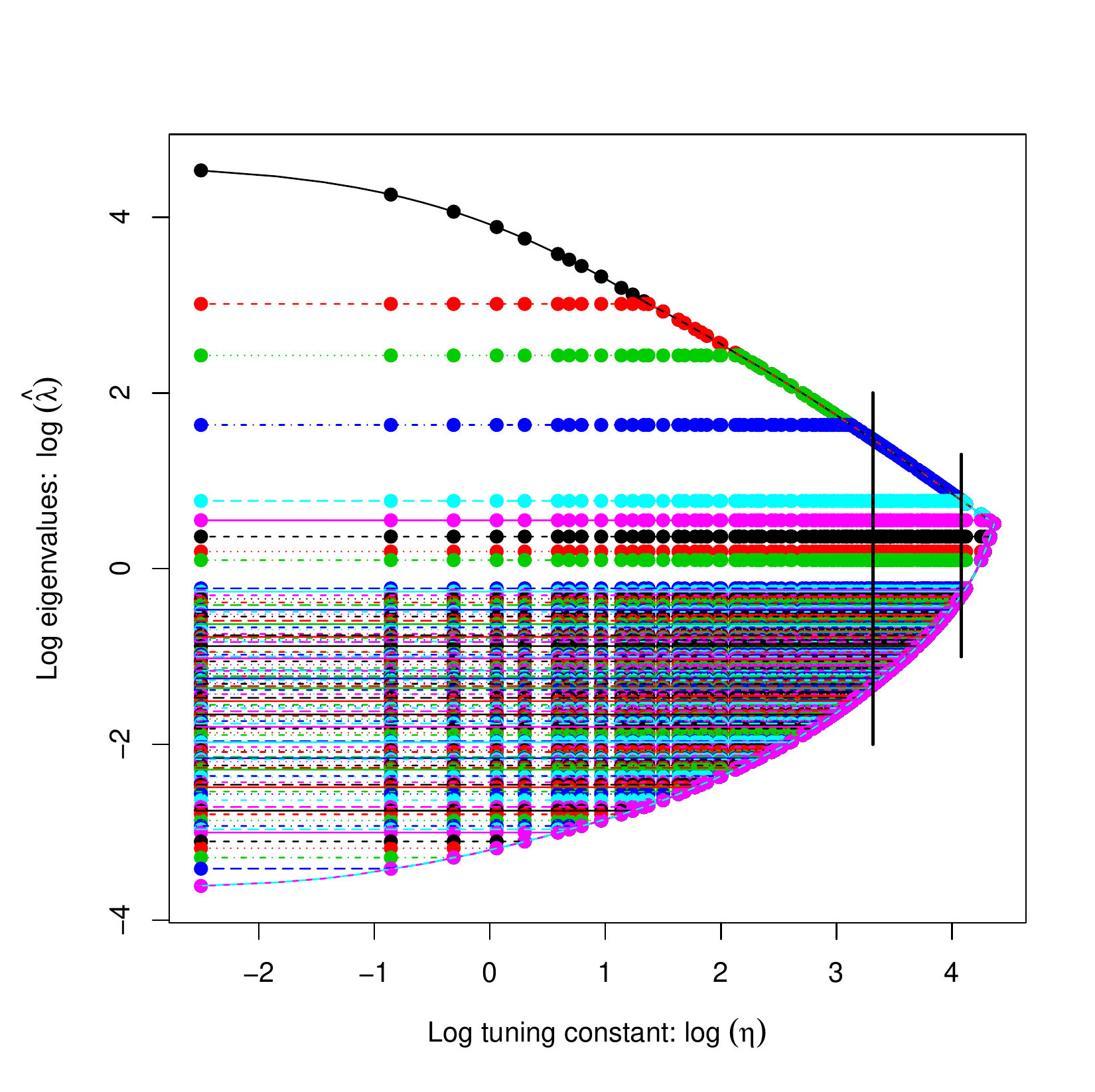}} 
\vspace*{-.6cm}
\end{center}
\caption{The elasso results for the call center data using the Mar\"{c}enko-Pastur weights (left) and the log condition number (right).
The results are plotted on a log-log scale. The first vertical line in each plot corresponds to the values of $\eta$ obtained via cross validation.
The second vertical line in each plot corresponds to the model obtained via model cross validation.
\label{fig5}}
\end{figure}

Both \citet{Huang-etal:2006} and \citet{Fan-etal:2009} give penalized covariance estimates based on a training set consisting of the
first $n = 205$ data points, using different types of penalties and tuning via five fold cross validation. We refer the reader to those 
papers for a discussion of the penalties used within. To make our analysis comparable to these
earlier estimates, we also consider only the first $n = 205$ data points and use five fold cross validation. Note that
an individual observation $x$ can be viewed as a nonstationary univariate time series of length 102. Rather than attempt to model this 
univariate time series, we use the elasso to try to achieve a parsimonious model for its covariance matrix. Figure \ref{fig5} show the 
results of the elasso when using the Mar\u{c}enko-Pastur weights and when using the weights associated with the log condition number. 

For the Mar\u{c}enko Pastur weights, the minimum 5-fold cross validation mean is $490.2$, with a standard error of $242.3$,
obtained at $\eta = 1.1$ ($\log(\eta) = 0.095$). 
By comparison, the cross validation mean for the sample covariance matrix is $5961.4$, and for the penalized estimate studied
in \citet{Huang-etal:2006} it is reported to be $3168.3$. For the elasso, the model at $\eta = 1.1$ corresponds to a single spiked
covariance model with the $19$ largest eigenvalues having multiplicity one, and the smallest eigenvalue having
multiplicity $83$. The minimum of the 5-fold model cross validation is $371.7$, which corresponds to partitioning of the eigenvalues into 
$10$ groups with the largest nine eigenvalues having multiplicity one, and the smallest eigenvalue having multiplicity $93$. 
If we use a relaxed lasso for this example, i.e.\ the largest values of $\eta$ for which the resulting cross validation mean is within 
one standard error of the value at $\eta = 1.1$, then one obtains a value of $\eta = 1.9$. This then corresponds to partitioning of 
the eigenvalues into $8$ groups with the largest even eigenvalues having multiplicity one, and the smallest eigenvalue having multiplicity $95$. 

For the log condition number penalty, the minimum 5-fold cross validation mean is $1457.8$, with a standard error of $291.9$ obtained 
at $\eta = 27.5$ ($\log(\eta) = 3.31$). 
This result gives a partitioning of the eigenvalues into $51$ groups, with the largest eigenvalue having multiplicity $4$, the smallest eigenvalue 
having multiplicity $53$, and the other $45$ eigenvalues having multiplicity one.  The minimum of the 5-fold model cross validation is $550.72$, 
which corresponds to the partitioning of the eigenvalues into $9$ groups, with the largest eigenvalue having multiplicity $4$, the smallest eigenvalue 
having multiplicity $91$, and the other $7$ eigenvalues having multiplicity one. The results based on the log condition number give less refined results 
and a worse fit than the results when using the Mar\u{c}enko Pastur weights. In general, the log condition number penalty does not allow for 
generating all possible partitions of the eigenvalues, only partitions of the form 
$\left\{ \{1, \ldots, r\}, \{r+1\}, \ldots, \{p\}, \{p+1, \ldots, q\} \right\}$,
and so as noted previously does not give consistent model paths in general. We recommend using the Mar\u{c}enko-Pastur
penalty over the log condition number for both the penalization problem and the dual constrained estimation problem.

By using the training set to fit the model, the remaining
$34$ observation can serve as a test set.  Consider partitioning the $q = 102$ dimensional 
response vector $x$ into a $p$-dimensional vector $x^{(1)}$ consisting of the first $p$ variables
and $x^{(2)}$ consisting of the other $p-q$ variables. For $p = 51$,
both \citet{Huang-etal:2006} and \citet{Fan-etal:2009} use $x^{(1)}$  to predict $x^{(2)}$ 
for the test set via the multivariate linear regression 
$\widehat{x}^{(2)} = \mu^{(2)} + \Sigma_{21}\Sigma_{11}^{-1}(x^{(1)}-\mu^{(1)})$. The
values of $\mu$ and $\Sigma$ are estimated from the training set using its sample mean 
and a penalized sample covariance respectively.

For $t = 52, \ldots, 102$, define the average absolute forecast error at component $t$ to be
\[ \mbox{AAFE}_t = \frac{1}{34}\sum_{i=206}^{239} |\widehat{x}_{it} - x_{it}|.\]
For the sample covariance matrix, the average AAFE is $1.46$. For various penalized covariance estimators, namely a LASSO, an adaptive LASSO, and SCAD, \citet{Fan-etal:2009} reports values of $1.39$, $1.34$ and 
$1.31$ respectively. The average AAFE for the elasso using the Mar\u{c}enko-Pastur weights at $\eta = 1.1$ is $1.35$. Curiously,
however, as $\eta$ increases in the elasso, the average AAFE monotonically decreases from $1.46$ to $1.19$. The last value corresponds to using $\widehat{\Sigma}= \overline{d}I$, i.e.\ predicting $x^{(2)}$ simply by its mean in the training set. 
This may not be unreasonable given that the 5-fold cross validation of $\widehat{\Sigma}= \overline{d}I$ is $787.5$, which is only $1.23$ standard errors larger than the smallest cross validation error obtained at $\eta = 1.1$.  
Overall, the average AAFE may not be the best measure of the overall performance of a covariance estimator, since it does not take into account all of the elements of $\Sigma$.  

\begin{figure}[h]
\begin{center}
\scalebox{.5}[.5]{
\includegraphics{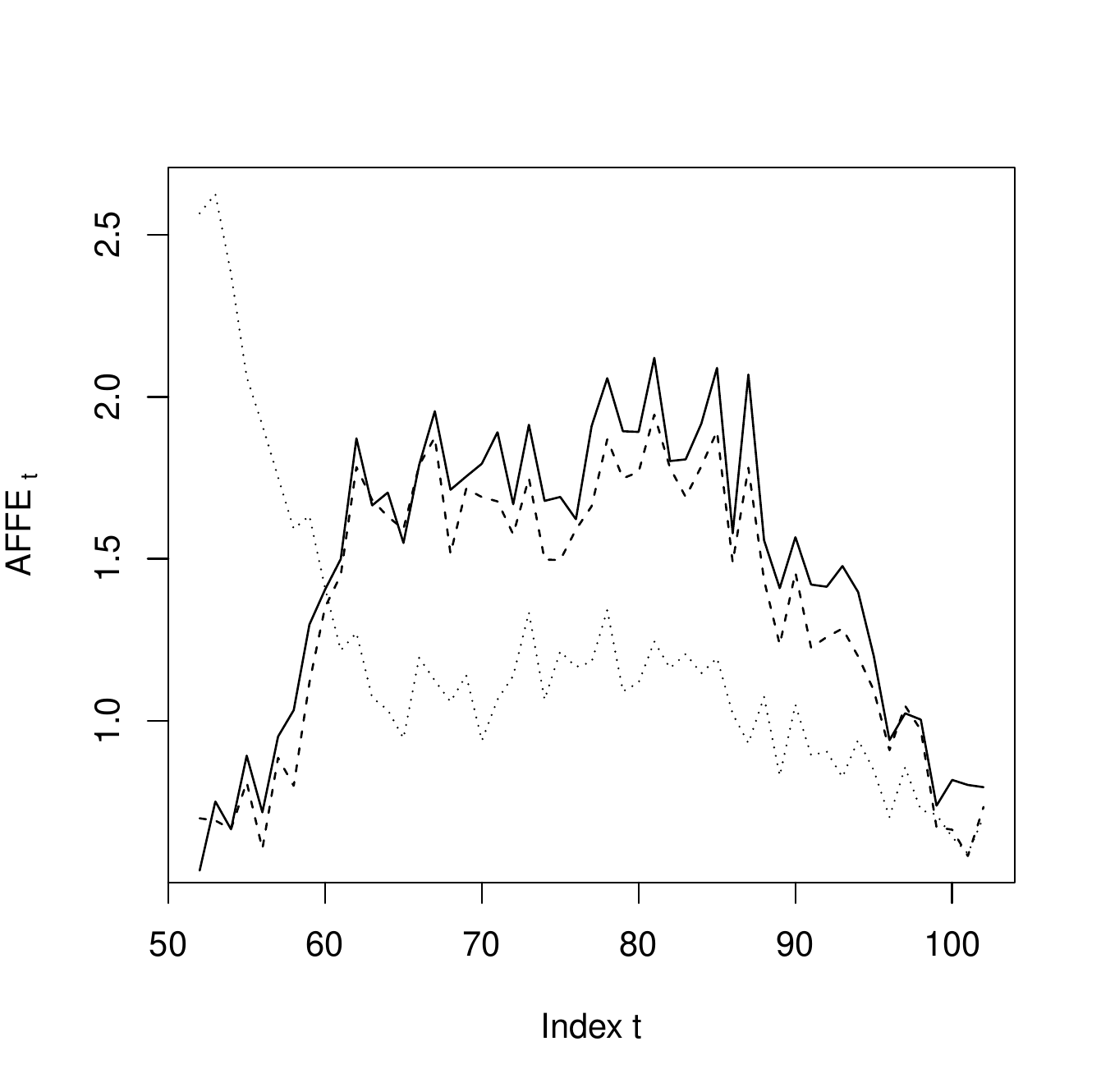}} 
\scalebox{.5}[.5]{
\hspace*{-.1cm}
\includegraphics{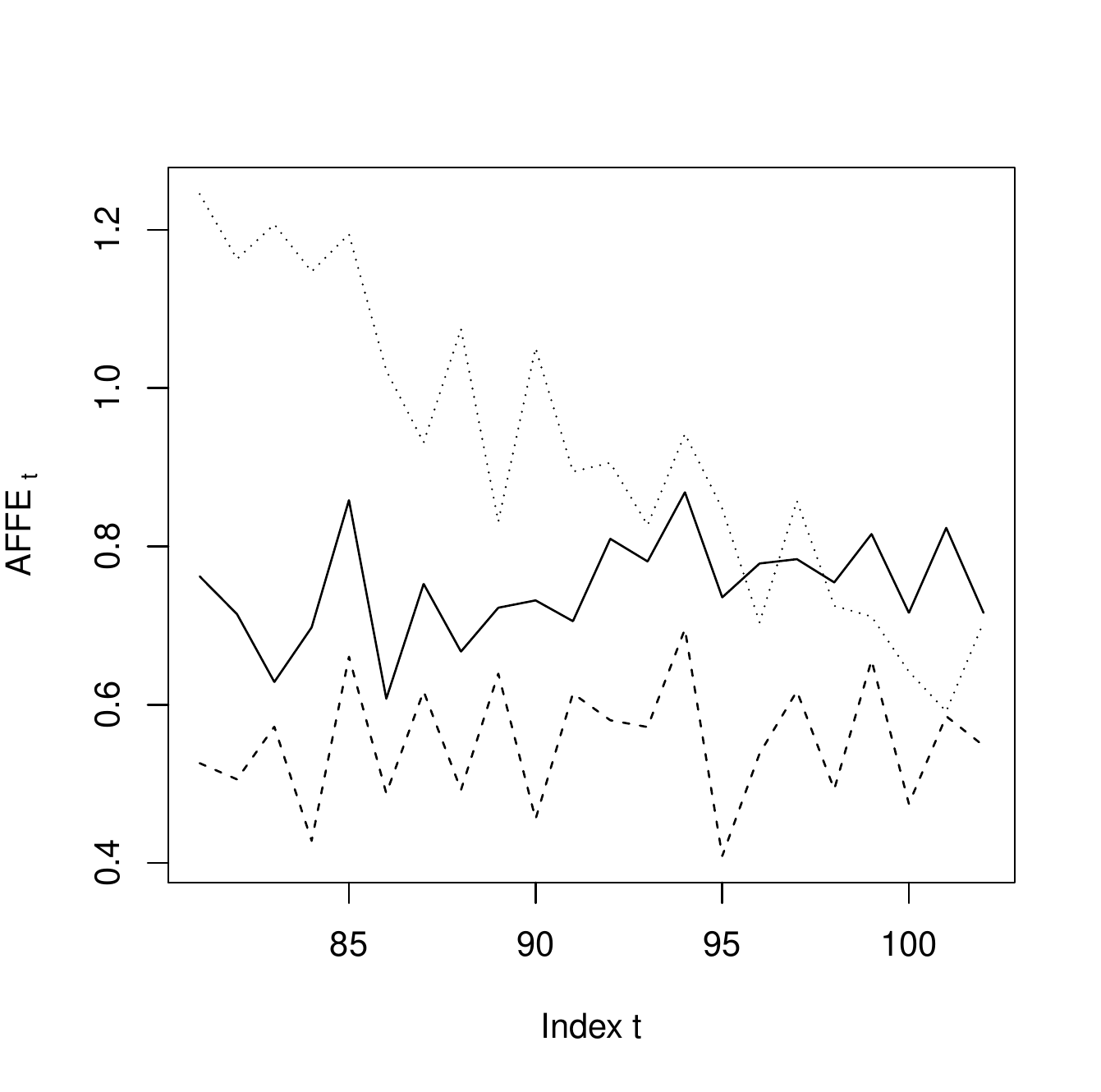}} 
\vspace*{-.6cm}
\end{center}
\caption{$\mbox{AAFE}_t$ based on the sample covariance matrix (solid line), the elasso estimate with minimal cross validation error (dashed line), and when using the mean of the training set as the 
predictor (dotted line). In the left plot, the first $51$ components are used to predict the last $51$ 
components. In the right plot, the first $80$ components are used to predict the last $22$ components.
\label{fig6}}
\end{figure}

To obtain a more detailed perspective, consider the plots in figure \ref{fig6}. The left plot shows graphs of $\mbox{AAFE}_t$ based on the following three estimates of $\Sigma$: the sample covariance matrix, the elasso estimate at $\eta = 1.1$, and $\overline{d}I$. The plot based on the elasso estimate is similar to the plot given in \citet{Huang-etal:2006} for their penalized estimator. It can be noted from the plot that simply using the mean of the training set as the predictor gives worse predictions 
for $t = 52, \ldots, 60$ but better predictions for $t = 61, \ldots, 102$. The right plot show the same graphs, but when using the first $p=80$ components to predict the last $q-p=22$ components. Note that
more elements of $\Sigma$ are involved in prediction when $p=80$ as oppose to when $p=51$. For $p=80$,
it can be seen that using the penalized covariance gives uniformly lower $\mbox{AAFE}_t$ compared to 
the other two methods.

\subsection{Discussion} \label{Sec:conclude}
The intent of this paper is to introduce the elasso method, as well as to give general results on penalized covariance matrices when using orthogonally
invariant penalties. Many open problems regarding the elasso still exist. Further study as to the choice of weights for the elasso method may be fruitful,
although we are fairly confident that the Mar\u{c}enko-Pastur weights is one of the best choices.  Other methods for choosing the
tuning constant in the elasso is worth exploring. In particular, one could use cross validation over a different criterion. For example, for the
call centre data, a cross validation method which measures the predictive ability of a subset of the variables for the other variables may be more
appropriate. 

Another possibility for tuning is to use an oracle method for minimizing the mean square error $E[\| \widehat{\Sigma}_\eta - \Sigma_o\|^2]$ 
\citep{Bickel-Levina:2008} or some other measure of the deviation between $\widehat{\Sigma}_\eta$ and its unknown target $\Sigma_o$ 
\citep{Chen-etal:2011,Ollila-Tyler:2014}. Under some models, the mean square error may be greatly reduced 
when using an elasso estimator in comparison to the sample covariance matrix. For example, when $\Sigma_o = \sigma^2 I$, a properly tuned elasso 
will give the estimate $\overline{d}I$ with high probability.  Under multivariate normality, and using the Frobenius norm, when $\Sigma_o = \sigma^2 I$ 
one obtains $E[\| S_n - \sigma^2 I\|^2] = q(q+1)\sigma^2/n$,  whereas $E[\| \overline{d}I - \sigma^2 I\|^2] = 2\sigma^2/n$. This reduction in
mean square error can be attributed to using a model with only one parameter for $\Sigma$ as opposed to $q(q+1)/2$ parameters.
In general, the number of parameters for the covariance model associated with a partitioning of the eigenvalues into $g \le q$ groups can be 
shown to be $q(q+1)/2 - m(m-1)/2 - (q-g)$, where $ m \le q - g + 1$ represent the cardinality of the largest group. Under the high-dimensional 
scenario $m/q \to \tau$ as $q \to \infty$, the proportional reduction in parameters converges to $\tau^2 \times 100\%$.

An important property of the elasso is that it generates a set of $q$ hierarchical models for the eigenvalue multiplicities. Rather than use
model cross validation to choose one of these $q$ models, another possibility would be to use sequential testing. That is, first consider the
model $\mG_1: \Sigma_o \propto I$, and perform a test for sphericity \citep{Anderson:2003, Muirhead:1982}. The classical test for 
sphericity is against the general alternative. Given the set of hierachical models $\mG_1, \ldots, \mG_q$, though, one could instead use the 
sequence of likelihood ratio tests for $\mG_1$ versus $\mG_2$, $\mG_2$ versus $\mG_3$, and so on. One drawback to such a testing approach is that the
null distributions tend only to be known asymptotically, for fixed $q$. Furthermore, as is the case for tests of subsphericity, i.e.\ testing if a
subset of the roots are equal \citep{Anderson:2003, Muirhead:1982}, the sample sizes needed for the asymptotic results to provide good approximations 
are inversely related to the separation of the eigenvalues in the models $\mG_k$ for $2 \le k \le q-1$. There has been some recent activity in 
developing asymptotic result for the test for sphericity in the large $q$, large $n$ setting \citep{Li-Yao:2016}, but as far as we are aware these 
results have not been extended to the more challenging case of testing for subsphericity.
From a pragmatic perspective, population eigenvalues may seldom be exactly equal. However, if the theoretical roots are distinct but 
not well separated enough to detect that they are distinct, then rather than focus on the individual eigenvectors, attention should be given to the 
joint eigenspaces associated with groups of eigenvalues which are not well separated. 

Finally, we note that $S_n$ can be replaced by any estimator of the covariance matrix, say $\widetilde{S}$. For example, $\widetilde{S}$ 
may be a more robust estimate of covariance matrix. In such cases, rather than associating $l(\Sigma;\widetilde{S})$ with the
negative log-likelihood function under multivariate normal sampling, one can view 
$l(\Sigma;\widetilde{S}) - \log\{\det(\widetilde{S})\} = \tr(\Sigma^{-1}\widetilde{S}) - \log\{\det(\Sigma^{-1}\widetilde{S})\}$ 
simply as a discrepancy measure between $\Sigma$ and $\widetilde{S}$, and then minimize its penalized version.

\section*{Acknowledgement}
We thank Jianhua Huang and Haipeng Shen for providing us with the edited version of the call centre data used in our paper.
The orginal data set was data set was made available to them by Avi Mandelbaum. An R-package to implement the elasso is currently being
developed together with Klaus Nordhausen. 

\appendix
\section*{Appendix: Proofs and some technical details} \label{App:proofs} 
\subsection*{Proofs for section \ref{Sec:SCM}}
\emph{Proof of Lemma \ref{Lem:orth}:}  Let $\Lambda = \diag\{\lambda_1, \ldots, \lambda_q\}$, and denote the spectral value decomposition of $\Sigma$ by
$\Sigma = Q\Lambda Q^\tran$. By orthogonal invariance, it follows
that $\Pi(\Sigma) = \Pi(\Lambda) = \Pi(R\Lambda R^\tran)$, with $R$ being a permutation matrix. Hence, the function $\pi(y_1, \ldots, y_q) = \Pi(e^\Delta)$,
with $\Delta = \diag\{y_1, \ldots, y_q\}$, is symmetric with $\Pi(\Sigma) = \pi(\log(\lambda_1), \ldots, \log(\lambda_q)$. \quad $\square$
\\

\noindent
\emph{Proof of Lemma \ref{Lem:diag}:} Again, express $\Sigma = Q\Lambda Q^\tran$. Also, let $H = [h_1 \cdots h_q] = P_n^\tran Q  \in \mathcal{O}_q$, and
define $\kappa_1 = \lambda_1^{-1}$ and $\kappa_j = \lambda_j^{-1} - \lambda_{j-1}^{-1}, j = 2, \ldots, q$. Since $\kappa_j  \ge 0$, the extremal properties
of eigenvalues gives
\[ \begin{array}{lcl}
 \tr\{\Sigma^{-1}S_n\} &=& \tr\{H\Lambda^{-1}H^\tran D_n\}  = \sum_{j=1}^q \lambda_j^{-1} h_j^\tran D_n h_j = 
\sum_{j=1}^q \kappa_j \left\{\sum_{k=j}^q h_k^\tran D_n h_k \right\} \\
 &\ge&  \sum_{j=1}^q \kappa_j \left\{\sum_{k=j}^q d_k \right\} = \sum_{j=1}^q d_j/\lambda_j = \tr\{\Lambda^{-1}D_n\}, 
\end{array} \]
with equality when $Q = P$. The lemma follows since $\det\{\Sigma\} = \det\{\Lambda\}$
and $\Pi(\Sigma) = \Pi(\Lambda)$. \quad $\square$ \\

\noindent
\emph{Proof of Theorem \ref{Thrm:nuniq}:} As previously noted, the first part of the lemma follows from Lemmas \ref{Lem:orth} and \ref{Lem:diag} and the strict 
convexity of $h(y;d,\eta)$. To prove continuity, we first state the following general lemma.
\begin{lemma} \label{Lem:cont}
Let $\mathcal{D}$ be a closed subset of $\R^p$. Suppose the real-valued functions $f(x)$ and $g(x)$ are continuous on $\mathcal{D}$, with $g(x) > 0$.
Furthermore, suppose $h(x;\eta) = f(x) + \eta g(x)$ has a unique minimum in $\mathcal{D}$ for any $0 \le \eta_o \le \eta \le \eta_1$.  If the set 
$\{x \in \mathcal{D} \ | \ h(x;\eta_o) \le c \}$ is compact for any $c \ge \inf \{h(x;\eta_o) \ | \ x \in \mathcal{D} \}$, then the function
$x(\eta) = \mbox{arginf}\{h(x;\eta) \ | \ x \in \mathcal{D} \}$ is continuous for $\eta_o \le \eta < \eta_1$.
\end{lemma}

To prove this lemma, first note that  $h(x;\eta)$ is increasing in $\eta$, and so the set $\{x(\eta) \ | \  \eta_o \le \eta < \eta_1 \}$ is
contained in the compact set $\{ x \ | \ h(x;\eta_o) \le h(x(\eta_1);\eta_1) \}$. So, if $\eta_k \to \eta$, then
$x(\eta_k)$ has a convergent subsequence, say  $x(\eta_{k'}) \to \tilde{x}$. By definition, 
$h(x(\eta_{k'});\eta_{k'}) \le h(x(\eta);\eta_{k'})$. By continuity, the left hand side converges to 
$h(\tilde{x};\eta)$ and the right hand side converges to $h(x(\eta);\eta)$. By uniqueness, this implies $\tilde{x} = x(\eta)$. Hence,
$x(\eta_k) \to x(\eta)$, which establishes Lemma \ref{Lem:cont}. 

This lemma applies to \eqref{eq:halpha}, for which $f(y) = \sum_{j=1}^p \{ d_j e^{-y_j} + y_j \}$ and $g(y) = \pi(y)$.
By convexity, both $f$ and $g$ are continuous. Also, the level sets of $h(y;d,\eta)$ are compact since $h(y;d,\eta) \to \infty$ as $\|y\| \to \infty$.
Hence $(\log\{\widehat{\lambda}_1\}, \ldots, \log\{\widehat{\lambda}_q\})$ is continuous for $\eta > 0$, which implies the continuity of $\widehat{\Lambda}^{n,\eta}$ 
and $\widehat{\Sigma}_{\eta}$  as functions of $\eta \ge 0$. \quad $\square$ \\

\noindent
\emph{Proof of Theorem \ref{Thrm:constrain}:}
Under the conditions of Theorem \ref{Thrm:nuniq}, $\kappa(\eta)$ is continuous and non-increasing with $\kappa_L \le \kappa(\eta) \le  \kappa_U$. 
Continuity follows since both $\Pi$ and $\widehat{\Sigma}_\eta$ are continuous. To prove that $\kappa(\eta)$ is non-increasing, 
suppose $\eta_1 < \eta_2$ and define, for $j = 1, 2$,
$\widehat{\Sigma}_j =\widehat{\Sigma}_{\eta_j}$, $l_j = l(\widehat{\Sigma}_j;S_n)$ and $\kappa_j = \kappa_{\eta_j}$. 
By definition of $\widehat{\Sigma}_j$, it follows that $l_1 + \eta_1 \kappa_1 \le l_2 + \eta_1 \kappa_2$ and $l_2 + \eta_2 \kappa_2 \le l_1 + \eta_2 \kappa_1$,
which together implies $\eta_1 (\kappa_1 - \kappa_2) \le l_2 - l_1 \le \eta_2 (\kappa_1 - \kappa_2)$. Hence $\kappa_1 \ge \kappa_2$.
Furthermore if $\kappa_1 = \kappa_2$, then $l_1 = l_2$ and
$l_1 + \eta_1 \kappa_1 = l_2 + \eta_1 \kappa_2$, which, by uniqueness, implies $\widehat{\Sigma}_1 = \widehat{\Sigma}_2$ .

Next, for $\kappa_L < \kappa \le \kappa_U$, define $\eta(\kappa) \equiv \inf\{ \eta \ge 0 ~|~ \kappa(\eta) = \kappa \}$, and note that
$\Pi(\widehat{\Sigma}_{\eta(\kappa)}) = \kappa$. It readily follows
from the previous paragraph that $\eta(\kappa)$ is strictly decreasing and continuous from above. By definition, 
$l(\widehat{\Sigma}_{\eta(\kappa)};S_n) + \eta(\kappa)\kappa \le l(\Sigma;S_n) + \eta(\kappa)\Pi(\Sigma)$ for any $\Sigma > 0$, and so 
$l(\widehat{\Sigma}_{\eta(\kappa)};S_n) - l(\Sigma;S_n) \le \eta(\kappa)(\Pi(\Sigma) - \kappa)$. Hence, if $\Pi(\Sigma) \le \kappa$, then
$l(\widehat{\Sigma}_{\eta(\kappa)};S_n) \le l(\Sigma;S_n)$, which implies $\widetilde{\Sigma}_\kappa =\widehat{\Sigma}_{\eta(\kappa)}$.
Since $\widehat{\Sigma}_{\eta}$ is continuous, it follows that $\widetilde{\Sigma}_\kappa$ is continuous at points of continuity of
$\eta(\kappa)$. It is also continuous at points of discontinuity of $\eta(\kappa)$ since $\widehat{\Sigma}_{\eta}$ is constant on
the sets $\{\eta ~|~ \kappa(\eta) = \kappa\}$. \quad $\square$
	
\subsection*{Proofs for section \ref{Sec:non-smooth}}
\emph{Proof of Lemma \ref{Lem:class}:} The functions $\pi_r(y) = \sum_{j=1}^q y_{(j)}$ are convex for $r=1, \ldots, q$, with $\pi_q(y) = \sum_{j=1}^r y_{j}$
being linear. The function $\pi(y)$ is symmetric and can be expressed as 
$\pi(y) = \sum_{r=1}^q b_r \pi_r(y)$ where $b_r = a_r - a_{r+1} \ge 0$ for $r = 1, \ldots, q-1$ and $b_q = a_q$. Each of the summands $b_r \pi_r(y)$
is convex, and hence $\pi(y)$ is convex. \quad $\square$ \\

\noindent
\emph{Proof of Lemma \ref{Lem:alphaG}:} 
The inequality $\widehat{\lambda}_k(\mathcal{G}) > \widehat{\lambda}_{k+1}(\mathcal{G}) $ holds if and only if
$\tilde{d}_k - \tilde{d}_{k+1} > \eta (\tilde{a}_k \tilde{d}_{k+1} - \tilde{a}_{k+1}\tilde{d_{k}})$,
which holds if and only if $\eta < \widehat{\eta}_k$. Hence, the inequality holds for all 
$k = 1, \ldots, r-1$ if and only if $\eta <  \eta(\mathcal{G})$. \quad $\square$
\\

\noindent
\emph{Proof of Theorem \ref{Thrm:nmin}:}
By definition \eqref{eq:Gr}, $\widehat{\lambda}_{k}(\mG_{r-1}) = \widehat{\lambda}_{k}(\mG_{r})$ for $k < k_r^*$ and 
$\widehat{\lambda}_{k}(\mG_{r-1}) = \widehat{\lambda}_{k+1}(\mG_{r})$ for $k > k_r^*$. Also, 
it can be shown that $\widehat{\lambda}_{k_r^*}(\mG_{r-1}) = \gamma \widehat{\lambda}_{k_r^*}(\mG_{r}) + (1-\gamma)\widehat{\lambda}_{k_r^*+1}(\mG_{r})$,
for some $0 < \gamma < 1$. Specifically, 
$\gamma =m_{k_r^*}(1+\eta \widetilde{a}_{k_r^*})/\{m_{k_r^*}(1+\eta \widetilde{a}_{k_r^*}) + m_{k_r^*+1}(1+\eta \widetilde{a}_{k_r^*+1})\}$, with
$\widetilde{a}_k$ and $m_k$ being defined with respect to the partition $\mG_r$. This implies that if 
$\widehat{\lambda}_{1}(\mG_{r}) > \cdots > \widehat{\lambda}_{r}(\mG_{r})$ then $\widehat{\lambda}_{1}(\mG_{r-1}) > \cdots > \widehat{\lambda}_{r-1}(\mG_{r-1})$.
By Lemma \ref{Lem:alphaG}, the former holds if and only if $\eta < \eta(\mG_r)$ and the latter holds if and only if $\eta < \eta(\mG_{r-1})$. Thus,
$\eta(\mG_r) < \eta(\mG_{r-1})$, with strict inequality holding since it is assumed that $k_r^*$ is well defined. 

As already noted, if $0 \le \eta < \eta(\mG_q)$ it ready follows that $\widehat{\lambda}_j = \widehat{\lambda}_j(\mG_q)$. 
We use finite induction to complete the proof. Suppose for $\eta(\mG_{r+2}) \le \eta < \eta(\mG_{r+1})$, we have 
$\widehat{\lambda}_j = \widehat{\lambda}_k(\mG_{r+1})$ for $j \in G_{r+1}(k)$. It then follows from the continuity of the solution in 
$\eta$, see Theorem \ref{Thrm:nuniq}, that for $\eta = \eta(\mG_{r+1})$ the solution corresponds to 
$\widehat{\lambda}_j = \widehat{\lambda}_k(\mG_{r})$ for $j \in G_{r}(k)$. This solution also holds for any $\eta(\mG_{r+1}) \le \eta < \eta(\mG_{r})$,
since otherwise if $\mG_r$ was not the optimizing partition for some $\eta$ in the interval, then there would be a discontinuity of the solution at
that value of $\eta$. \quad $\square$ 

\subsection*{Technical details for sections \ref{Sec:CV} and \ref{Sec:discuss}}
Although the conditions in Theorem \ref{Thrm:nmin} that the eigenvalues of $S_n$ be distinct and that $k_r^*$ be unique hold 
with probability one when random sampling from a continuous multivariate distribution, they are not necessary. For the penalty $\Pi(\Sigma;a)$,
consider the general problem of minimizing $L(\Sigma;\tilde{S},\eta)$ over $\Sigma > 0$, where $\tilde{S} > 0$ is some given
matrix, e.g.\ the population covariance matrix.  For this general case, the above conditions on $\tilde{S}$ may not hold.  
Theorem \ref{Thrm:nmin} then requires a slight modification, namely $0 \le \eta(\mG_q) \le \cdots \le \eta(\mG_2) < \eta(\mG_1) = \infty$. That is, 
the knots of the elasso are not necessarily unique.  With this modification, the statement of \ref{Thrm:nmin} holds. 

If the eigenvalues of $\tilde{S}$ lie in $p < q$ distinct groups, then $0 = \eta(\mG_q) = \cdots = \eta(\mG_{p+1}) < \eta(\mG_p)$.
For example, if $\tilde{S} \propto I$, then $0 = \eta(\mG_q) = \cdots = \eta(\mG_{2}) < \eta(\mG_1) = \infty$.
In general, if $k_r^*$ is not unique, but rather the infimum in its definition \eqref{eq:Gr} is obtain at $t \le r-1$ points, then $t$ knots occur at
the same point, namely $\eta(\mG_r) = \cdots = \eta(\mG_{r-t+1})$. 

The above results can be applied to generating the elasso for a given multi-spike model. Consider minimizing $L(\Sigma;S_n,\eta)$ over
all $\Sigma > 0$ for which the multiplicities of the ordered eigenvalues are $m_1, \ldots, m_p$ respectively, with $m_1 + \cdots + m_p = q$ and $p < q$.
Let $\mG_o = \{G_o(1), \ldots, G_o(p)\}$ denote the corresponding grouping of the eigenvalues, and let $\tilde{d}_k$ denote the average of the
eigenvalues of $S_n$ in the group $G_o(k)$, for $k = 1, \ldots, p$.
It can be shown that the solution to this problem is then the same as the solution to the problem of 
minimizing $L(\Sigma;\tilde{S},\eta)$ over $\Sigma > 0$ where $\tilde{S}$ is the maximum likelihood estimate of $\Sigma$ under the model.
That is, $\tilde{S} = P_n\tilde{D}P_n^\tran$, with $\tilde{D}$ being a diagonal matrix with elements $\tilde{d}_k$ repeated $m_k$ times, 
for $k = 1, \ldots, p$. When random sampling from a continuous multivariate distribution, $k_r^*$ is unique with probability one for
$r \le p$ and hence there are $p$ distinct knots $0 = \eta(\mG_q) = \cdots = \eta(\mG_{p+1}) < \eta(\mG_p) < \cdots < \eta(\mG_2) < \infty $.

\subsection*{Proofs for section \ref{Sec:fixed}}
\emph{Proof of Lemma \ref{Lem:asym1}:}
Let $\Sigma_\eta$ be the unique minimum of $L(\Sigma;\Sigma_o,\eta)$ over $\Sigma_o$, or in other words, $\Sigma_\eta$ 
is the population or functional version of $\widehat{\Sigma}_\eta$. Also, without loss of generality, assume there 
exists a $\Sigma_*$ such that $\Pi(\Sigma_*) = 0$.

\emph{Part (a)}: Consider a point in the sample space such that $S_n \to \Sigma_o$. If $\lambda_1(\widehat{\Sigma}_\eta) \to \infty$ or 
$\lambda_q(\widehat{\Sigma}_\eta) \to 0$, with $\eta$ possibly depending on $n$, then it follows that $l(\widehat{\Sigma}_\eta;S_n) \to \infty$. 
This implies $l(\Sigma_*;S_n) = L(\Sigma_*;S_n,\eta) \ge L(\widehat{\Sigma}_\eta;S_n,\eta) \to \infty$, which is a contradiction since 
$l(\Sigma_*;S_n) \to l(\Sigma_*,\Sigma_o)$. Hence, 
\begin{equation} \label{eq:compact}  \{ \widehat{\Sigma}_{\eta} \ | \ \eta \ge 0, \ n = 1, 2, ... \}  \ \mbox{is contained in some compact set.}
\end{equation}
Suppose $\eta \to \eta^o$ as $n \to \infty$, then by \eqref{eq:compact} it follows there exist a convergent sub-sequence, say
$\widehat{\Sigma}_{\eta} \to \widetilde{\Sigma}_{\eta^o}$. By definition $L(\Sigma_\eta;\Sigma_o,\eta) \le L(\widehat{\Sigma}_\eta;\Sigma_o,\eta)$,
and $L(\Sigma_\eta;S_n,\eta) \ge L(\widehat{\Sigma}_\eta;S_n,\eta)$. Since $L(\Sigma;S,\eta)$ is continuous in all three arguments, taking limits give
$L(\Sigma_{\eta^o};\Sigma_o,\eta_o) \le L(\widetilde{\Sigma}_{\eta^o};\Sigma_o,\eta_o)$,
and $L(\Sigma_{\eta^o};\Sigma_o,\eta) \ge L(\widetilde{\Sigma}_{\eta^o};\Sigma_o,\eta_o)$. So,
$L(\Sigma_{\eta^o};\Sigma_o,\eta) = L(\widetilde{\Sigma}_{\eta^o};\Sigma_o,\eta_o)$. By uniqueness, this implies $\widetilde{\Sigma}_{\eta^o} = \Sigma_{\eta^o}$.
Hence, since this holds for any sub-sequence and $S_n \to \Sigma_o$ almost surely, we have
\begin{equation} \label{eq:consistency}
\mbox{if} \ \eta \to \eta^o \ \mbox{almost surely, then} \ \widehat{\Sigma}_{\eta} \to \Sigma_{\eta^o} \ \mbox{almost surely}.
\end{equation}
Part (a) then follows as a special case of \eqref{eq:consistency} after noting $\Sigma_0 = \Sigma_o.$

\emph{Part (b)}: For $k = 1, \ldots, K$, let $\overline{x}_{k,n}$, $S_{k,n}$ and $\widehat{\Sigma}_{k,\eta}$ denote the sample mean vector, the sample
covariance matrix and the penalized estimate respectively computed from the data not in $\mA_k$.  Observe that $\widehat{\Sigma}_{k,0} = S_{k,n}$. 
Also, for $\mA = \mA_k$, let $S^*_{k,n} = n_{\mA}^{-1} \sum_{x_i \in \mA} (x_i - \overline{x}_{k,n})(x_i - \overline{x}_{k,n})^\tran$, and note that
$cv(\eta; \mA) = n_{\mA}~l(\widehat{\Sigma}_{k,\eta};S^*_{k,n})$. 

Consider a point in the sample space so that $\overline{x}_{k,n} \to \mu_o$ and $S^*_{k,n} \to \Sigma_o$ for $k = 1, \ldots, K$, which by the
strong law of large numbers occurs almost surely. This also implies $S_{k,n} \to \Sigma_o$ for $k = 1, \ldots, K$, and $S_n \to \Sigma_o$.
By compactness, i.e.\ by applying \eqref{eq:compact} to $\widehat{\Sigma}_{k,\eta}$, 
there exist convergent sub-sequences, say $\widehat{\Sigma}_{k,\eta^{cv}} \to \Sigma_k$ for $k = 1, \ldots K$.  By definition, 
$\sum_{k=1}^K  l(\widehat{\Sigma}_{k,\eta^{cv}};S^*_{k,n}) \le \sum_{k=1}^K  l(\widehat{\Sigma}_{k,\eta};S^*_{k,n})$, which by taking the limits on both sides
gives $\sum_{k=1}^K  l(\Sigma_k;\Sigma_o) \le K~l(\Sigma_o;\Sigma_o)$. However, since $l(\Sigma_o;\Sigma_o) \le l(\Sigma_k;\Sigma_o)$, this
implies $\sum_{k=1}^K  l(\Sigma_k;\Sigma_o) = K~l(\Sigma_o;\Sigma_o)$, which only holds if $\Sigma_k = \Sigma_o$ for $k = 1, \ldots, K$. 
Since this holds for any convergent sub-sequences, we have $\widehat{\Sigma}_{k,\eta^{cv}} \to \Sigma_o$ for $k = 1, \ldots K$

It still needs to be shown that $\widehat{\Sigma}_{\eta^{cv}} \to \Sigma_o$. To do so, two cases are considered. The first case is when $\eta^{cv}$ is bounded.
For this case, consider a sub-sequence such that $\eta^{cv} \to \eta < \infty$. From \eqref{eq:consistency}, it follows that 
$\widehat{\Sigma}_{\eta^{cv}} \to \Sigma_\eta$ and $\widehat{\Sigma}_{k,\eta^{cv}} \to \Sigma_\eta$. However, it has already been shown that
$\widehat{\Sigma}_{k,\eta^{cv}} \to \Sigma_o$ and hence $\Sigma_\eta = \Sigma_o$. Consequently, $\widehat{\Sigma}_{\eta^{cv}} \to \Sigma_o$.

The second case is when $\eta^{cv}$ is not bounded above. For this case, by \eqref{eq:consistency}, there exist a sub-sequence,  
say $\widehat{\Sigma}_{\eta^{cv}} \to \Sigma_{*,o}$, and such that $\eta^{cv} \to \infty$.
By definition, $l(\widehat{\Sigma}_{\eta^{cv}};S_{n})+\eta^{cv}\Pi(\widehat{\Sigma}_{\eta^{cv}}) \le l(\Sigma_*;S_n)$,
where $\Pi(\Sigma_*) = 0$. For this sub-sequence, $\Pi(\widehat{\Sigma}_{\eta^{cv}}) \to 0$, otherwise we have a contradiction, and so
$\Pi(\Sigma_{*,o}) = 0$. An analogous argument also gives $\Pi(\widehat{\Sigma}_{k,\eta^{cv}}) \to 0$, and since 
$\widehat{\Sigma}_{k,\eta^{cv}} \to \Sigma_o$, we have $\Pi(\Sigma_o) = 0$.
Finally, by definition, $l(\widehat{\Sigma}_{\eta_{cv}};S_n) \le L(\widehat{\Sigma}_{\eta_{cv}};S_n;\eta^{cv}) \le l(\Sigma_o;S_n)$. 
Passing to the limit gives $l(\Sigma_{*,o};\Sigma_o) \le l(\Sigma_o;\Sigma_o)$. The reverse inequality also holds, and so $\Sigma_{*,o} = \Sigma_o$.
Since this holds for any convergent sub-sequence, we have $\widehat{\Sigma}_{\eta^{cv}} \to \Sigma_o$.  \quad $\square$ \\

\noindent
\emph{Proof of Lemma \ref{Lem:asymm}:} Suppose $\Sigma_o$ has order eigenvalues with multiplicities $m_1, \ldots, m_p$ respectively,
where $m_1 + \cdots + m_p = q$. Let $\mG_o = \{G_o(1), \ldots, G_o(p)\}$ denote the corresponding partition. So, 
if $\mG_o \in \{\mG_1, \ldots, \mG_q\}$ then $\mG_o = \mG_p$.  It is then to be shown that $P( \mG_o = \mG_p) \to 1$.
The last statement implies convergence in probability. A stronger statement which is shown in this proof is convergence almost
surely, i.e.\ $P(\mG_o = \mG_p \ \mbox{as} \ n \to \infty) = 1$.

Consider a point in the sample space such that $S_n \to \Sigma_o$.
Let $\eta_o$ denote the first non-zero knot of the population elasso path $\Sigma_\eta$, as defined in 
the proof of Lemma \ref{Lem:asym1}, and consider $\eta < \eta_o$. The grouping of the
eigenvalues of $\Sigma_\eta$ are still given by $\mG_o$. Let $\widehat{\mG}_\eta$ denote the grouping of the eigenvalues of $\widehat{\Sigma}_\eta$.
By strong consistency \eqref{eq:consistency}, for large enough $n$,  $\widehat{\mG}_\eta \in \{\mG_p, \ldots, \mG_q\}$, with none the subsets within $\widehat{\mG}_\eta$ 
containing both an element from $G_o(j)$ and an element from $G_o(k)$ for $j \ne k$. That is, the groupings in $\mG_o$ correspond to unions of the groupings in $\widehat{\mG}_\eta$.
The proof can be completed then by showing the knot $\eta(\mG_{p-1}) < \eta_o$ for large enough $n$. From its definition in
Lemma \ref{Lem:alphaG}, though, it readily follows that $\eta(\mG_{p-1}) \to 0$ since $\tilde{a}_k > \tilde{a}_{k+1}$ by assumption. So, for large enough
$n$, the grouping $\mG_o$ occurs before $\eta_o$.  \quad $\square$ 

\bibliographystyle{agsm}

\begin{thebibliography}{9}


\bibitem[{Anderson(1965)}]{Anderson:1965}
\textsc{Anderson, G.~A.} (1965).
\newblock{An asymptotic expansion for the distribution of the latent roots of the estimated covariance matrix.}
\newblock \textit{Ann. Math. Statist.}, 1153 - 1173.

\bibitem[{Anderson(2003)}]{Anderson:2003}
\textsc{Anderson, T.~W.} (2003). 
\newblock \textit{{An introduction to multivariate statistical analysis.}}
\newblock New York: Wiley

\bibitem[{Bai-Yao(2012)}]{Bai-Yao:2012}
\textsc{Bai, Z.}  \& \textsc{Yao, J.} (2012).
\newblock{On sample eigenvalues in a generalized spiked population model}.
\newblock \textit{J. Mult. Anal.} {\bf 106}, 167 - 177.

\bibitem[{Baik-Silverstein(2006)}]{Baik:2006}
\textsc{Baik, J.} \& \textsc{Silverstein, J.~W.} (2006).
\newblock{Eigenvalues of large sample covariance matrices of spiked
population models}. 
\newblock \textit{J. Mult. Anal.} {\bf 97}, 1382 - 1408.

\bibitem[{Bickel-Levina(2008)}]{Bickel-Levina:2008}
\textsc{Bickel, P.~J.} \& \textsc{Levina, E.} (2008).
\newblock{Regularized estimation of large covariance matrices}.
\newblock \textit{Ann. Statist.}, 199 - 227.

\bibitem[{Chen, et al.(2011)}]{Chen-etal:2011}
\textsc{Chen, Y.},\textsc{Wiesel, A.} \& \textsc{Hero, A.~O.} (2011). 
\newblock{Robust shrinkage estimation of high-dimensional covariance matrices}.
\newblock \textit{IEEE Transactions on Signal Processing} {\bf 59}, 4097 – 4107.

\bibitem[{Davis, et al.(2014)}]{Davis-etal:2014}
\textsc{Davis, R.~A.}, \textsc{Zang, P.} \& \textsc{Zheng, T.} (2014).
\newblock{Reduced-rank covariance estimation in vector autoregressive modeling}.
\newblock \textit{arXiv preprint arXiv:}, 1412.2183.

\bibitem[{Fan, et al.(2009)}]{Fan-etal:2009}
\textsc{Fan, J.}, \textsc{Feng, Y.} \& \textsc{Wu,Y.} (2009).
\newblock{Network exploration via the adaptive lasso and scad penalties}.
\newblock \textit{Ann. Appl. Stat.}, {\bf 3}, 521 – 541.

\bibitem[{Friedman-Hastie(2008)}]{Friedman-Hastie:2008}
\textsc{Friedman, J.}, \textsc{Hastie, T.} \& \textsc{Tibshirani, R.} (2008).
\newblock{Sparse inverse covariance estimation with the graphical lasso}. 
\newblock \textit{Biostatistics} {\bf 9}, 432 - 441.

\bibitem[{Haff(1980)}]{Haff:1980}
\textsc{Haff, L.~R.} (1980).
\newblock{Empirical Bayes estimation of the multivariate normal covariance matrix}. 
\newblock \textit{Ann. Statist.}, 586 - 597.

\bibitem[{Haff(1991)}]{Haff:1991}
\textsc{Haff, L.~R.} (1991).  
\newblock{The variational form of certain Bayes estimators}.
\newblock \textit{Ann. Statist.}, 1163 - 1190.

\bibitem[{Huang, et al.(2006)}]{Huang-etal:2006}
\textsc{Huang, J.~Z.}, \textsc{Liu, N.}, \textsc{Pourahmadi, M.} \& \textsc{Liu, L.} (2006).
\newblock{Covariance matrix selection and estimation via penalised normal likelihood}.
\newblock \textit{Biometrika} {\bf 93}, 85 - 98.

\bibitem[{Johnstone(2001)}]{Johnstone:2001}
\textsc{Johnstone, I.~M.} (2001).
\newblock{On the distribution of the largest eigenvalue in principal components analysis}.
\newblock \textit{ Ann. Statist.}, 295 - 327.

\bibitem[{Ledoit-Wolf(2004)}]{Ledoit-Wolf:2004}
\textsc{Ledoit, O.} \& \textsc{Wolf, M.} (2004).
\newblock{A well-conditioned estimator for large-dimensional covariance matrices}.
\newblock \textit{J. Mult. Anal.} {\bf 88}, 365 - 411.

\bibitem[{Li-Yao(2014)}]{Li-Yao:2016}
\textsc{Li, Z.} \& \textsc{Yao, J.} (2016).
\newblock{Testing the sphericity of a covariance matrix when
the dimension is much larger than the sample size}.
\newblock \textit{arXiv preprint arXiv:} 1508.02498.

\bibitem[{Mar\u{c}enko-Pastur(1967)}]{MP:67}
\textsc{Mar\u{c}henko, V.~A.} \& \textsc{Pastur, L.~A.} (1967). 
\newblock{Distribution of eigenvalues for some sets of random matrices}.
\newblock \textit{Sbornik: Mathematics} {\bf 1}, 457 - 483.

\bibitem[{Meinshausen(2007)}]{Mein:07}
\textsc{Meinshausen, N.} (2007).
\newblock{Relaxed lasso}.
\newblock \textit{Computational Statistics \& Data Analysis} {\bf 52}, 374 - 393.

\bibitem[{Mestre(2008)}]{Mestre:2008}
\textsc{Mestre, X.} (2008). 
\newblock{Improved estimation of eigenvalues and eigenvectors of covariance matrices using their sample estimates}.
\newblock \textit{IEEE Transactions on Information Theory} {\bf 54}, 5113 - 5129.

\bibitem[{Muirhead(1982)}]{Muirhead:1982}
\textsc{Muirhead, R.~J.} (1982).  
\newblock \textit{Aspects of multivariate statistical theory}. 
\newblock John Wiley \& Sons

\bibitem[{Ollila-Tyler(2014)}]{Ollila-Tyler:2014}
\textsc{Ollila, E.} \& \textsc{Tyler, D.~E.} (2014). 
\newblock{Regularized M-estimators of scatter}.
\newblock \textit{IEEE Transactions on Signal Processing} {\bf 62}, 6059 - 6070.

\bibitem[{Paul(2007)}]{Paul:2007}
\textsc{Paul, D.} (2007). 
\newblock{Asymptotics of sample eigenstructure for a large dimensional spiked covariance matrix}.
\newblock \textit{Statistica Sinica} {\bf 17}, 1617 - 1642.

\bibitem[{Schott(2006)}]{Schott:2006}
\textsc{Schott, J.~R.} (2006).   
\newblock{A high-dimensional test for the equality of the smallest eigenvalues of a covariance matrix}. 
\newblock \textit{J. Mult. Anal.} {\bf 97}, 827 - 843.

\bibitem[{Stein(1975)}]{Stein:1975}
\textsc{Stein, C.} (1975).
\newblock \textit{{Estimation of a covariance matrix, Rietz Lecture}}.
\newblock 39th Annual Meeting IMS, Atlanta, GA 

\bibitem[{Stone(1974)}]{Stone:1974} 
\textsc{Stone, M.} (1974).
\newblock{Cross-validatory choice and assessment of statistical predictions}.
\newblock \textit{J.\ Roy.\ Statist.\ Soc.\ Ser.\ B.} {\bf 36}, 111 - 147.

\bibitem[{Warton(2008)}]{Warton:2008}
\textsc{Warton, D.~I.} (2008).  
\newblock{Penalized normal likelihood and ridge regularization of correlation and covariance matrices}.
\newblock \textit{J. Amer. Statist. Assoc.} {\bf 103}, 340 - 349.

\bibitem[{Wiesel(2012)}]{Wiesel:2012}
\textsc{Wiesel, A.} (2012).  
\newblock{Unified framework to regularized covariance estimation in scaled Gaussian models}.
\newblock \textit{IEEE Transactions on Signal Processing} {\bf 60}, 29 - 38.


\bibitem[{Won-etal(2013)}]{Won-etal:2013}
\textsc{Won, J.}, \textsc{Lim, J.}, \textsc{Kim, S.} \& \textsc{Rajaratnam, B.} (2013).  
\newblock{Condition number regularized covariance estimation}.
\newblock \textit{J.\ Roy.\ Statist.\ Soc.\ Ser.\ B.} {\bf 75}, 427 - 450.

\bibitem[{Yang-Berger(1994)}]{Yang-Berger:1994}
\textsc{Yang, R.} \& \textsc{Berger, J.~O.} (1994).
\newblock{Estimation of a covariance matrix using the reference prior}. 
\newblock \textit{Ann. Statist.}, 1195 - 1211.

\end{thebibliography}

\end{document}